\documentclass[%
 reprint,
 superscriptaddress,
 amsmath,amssymb,
 aps,
prb, citeautoscript,
]{revtex4-1}
\usepackage[version=3]{mhchem} 

\usepackage{graphicx}
\usepackage{dcolumn}
\usepackage{bm}
\usepackage{color, soul}
\usepackage{natbib}
\makeatletter
\newcommand*{\rom}[1]{\expandafter\@slowromancap\romannumeral #1@}
\usepackage[colorlinks, linkcolor=blue, citecolor=blue, urlcolor=blue]{hyper ref}
\makeatother
\usepackage{soul}

\begin{document}
\preprint{APS/123-QED}

\title{Defect formation in CsSnI$_3$ from Density Functional Theory and Machine Learning}

\author{Chadawan Khamdang}
\affiliation{Department of Electrical and Computer Engineering, State University of New York at Binghamton, Binghamton, New York 13902, United States}
\author{Mengen Wang}
\email{mengenwang@binghamton.edu}
\affiliation{Department of Electrical and Computer Engineering, State University of New York at Binghamton, Binghamton, New York 13902, United States}
\affiliation{Materials Science and Engineering Program, State University of New York at Binghamton, Binghamton, New York 13902, United States}

\begin{abstract}

Sn-based perovskites as low-toxicity materials are actively studied for optoelectronic applications. 
However, their performance is limited by $p$-type self-doping, which can be suppressed by substitutional doping on the cation sites.
In this study, we combine density functional theory (DFT) calculations with machine learning (ML) to develop a predictive model and identify the key descriptors affecting formation energy and charge transition levels of the substitutional dopants in CsSnI$_{3}$.
Our DFT calculations create a dataset of formation energies and charge transition levels and show that Y, Sc, Al, Zr, Nb, Ba, and Sr are effective dopants that pin the Fermi level higher in the band gap, suppressing the $p$-type self-doping.
We explore ML algorithms and propose training a random forest regression model to predict the defect formation properties.
This work shows the predictive capability of combining DFT with machine learning and provides insights into the important features that determine the defect formation energetics.

\end{abstract}

\maketitle
\section{\label{intro}Introduction}

Halide perovskites are promising candidates for optoelectronic applications due to their straightforward synthesis methods and optical and charge transport properties.\cite{1-Chemica-stability, 2-Electron-hole-diffusion-lengths-solution-grown-CH3NH3PbI3-prop1, 3-Electron-hole-diffusion-lengths-exceeding-prop2, 4-Chemical-management-for-colorful-efficient-and-stable-inorganic-organic-hybrid-nanostructured-solar-cells-prop3} 
The power conversion efficiency (PCE) of Pb-based perovskite-based solar cells (PSCs) has dramatically improved. \cite{5-pce3.8, 6-pce25} 
CsSnI$_3$ has been explored as a promising low-toxicity alternative to Pb-based perovskites.\cite{10-usesn2-All-solid-state, 11-usesn3-CsSnI3:semiconductorormetal?, 7-toxicpb1-The-importance-of-moisture-in-hybrid-lead-halide-perovskite, 8-toxicpb2-Toxicity-of-organometal-halide-perovskite} 
Despite its potential, the PCE of CsSnI$_3$ remains lower (14.8\%)\cite{12-pcesn} than that of CsPbI$_3$. 
This reduced efficiency is primarily attributed to the substantial self-$p$-type doping and defect-assisted nonradiative recombination.\cite{13-nonrad-Importance-of-reducing-vapor-atmosphere, 11-usesn3-CsSnI3:semiconductorormetal?, 14-ptype2-Influence-of-defects-and-synthesis, 15-ptype3-Effects-of-organic-cations}

To address these limitations, defect engineering through doping has been investigated as a potential solution to improve Sn-based perovskite properties. 
Experimental studies on Ba-doped Sn-Pb perovskites indicate that Ba incorporation can reduce hole concentration, thereby reducing the effects of $p$-type doping.\cite{16-yu2022gradient-Gradient-Doping-in-Sn--Pb-Perovskites} 
Density functional theory (DFT) calculations provide a theoretical understanding of the mechanism, showing that Ba acts as an energetically favorable donor in CsSnI$_3$ that shifts the Fermi level upward and decreases the background hole concentration.\cite{17-Simultaneous-suppression-of-p-doping-dft-bacs} 
DFT studies also propose that trivalent cation doping on the Sn site in MASnI$_3$ including Sc, La, and Ce can also raise the Fermi level, which is supported by experimental validation that La doping in MASnI$_3$ results in an increase in photocurrent and open circuit voltage.\cite{gregori2024trivalent_doping_MASnI3}
Another DFT study on MASnI$_3$/MASnI$_2$Br proposes that Sc, Y, and La doping can shift the Fermi level upward, thereby reducing hole concentration compared to pristine perovskites.\cite{gregori2024trivalent_doping_MASnIBr}

DFT is widely used to predict defect formation energies under various chemical potentials and has reliably predicted intrinsic defect and dopant formation energies and charge transition levels in semiconductors.\cite{14-ptype2-Influence-of-defects-and-synthesis, kang2021review_dft_perovskite, mu2022dopedAlGaO, 34-freysoldt2014first-First-principles-calculations-for-point-defects-in-solids} 
Defect calculations require large supercells and hybrid functionals with spin-orbit coupling (SOC) to correctly describe the electronic structure and charge localization, which are computationally demanding.\cite{24-socIodineinterstitials, 25-zhang2022defect, kang2021review_dft_perovskite, 35-zhang2022origins-Originsofp-Dopingincssni3}
To overcome these limitations, machine learning (ML) algorithms offer a promising approach to predict and understand defect properties efficiently. 
Recent studies have demonstrated that DFT can be combined with ML algorithms to predict formation energies and charge transition levels for both dopants and intrinsic defects.\cite{26-predictml3predictabo3, 25-predictml2impuritiesinhalideperovskites,24-predictmlzbuniversalmachinelearning,varley2017descriptor} 
Specifically for dopant incorporation energetics, data generated from DFT calculations using the PBE functional has been used to train ML algorithms to predict defect formation energies in perovskite oxides (ABO$_3$) and halide perovskites (MAPbX$_3$).\cite{26-predictml3predictabo3, 25-predictml2impuritiesinhalideperovskites} 
There is also a growing interest in applying ML algorithms to predict defect energetics at the hybrid functional accuracy.\cite{varley2017descriptor, 24-predictmlzbuniversalmachinelearning}
These studies reveal opportunities and the need to improve the prediction of defect formation energetics by combining DFT calculations with hybrid functionals and machine learning methods, which is also promising to provide insights into the physical and chemical descriptors underlying these properties.

This work combines DFT using HSE06+SOC with ML to predict formation energies and charge transition levels for substitutional dopants in CsSnI$_3$. 
We explore elements from group II-A (e.g., Mg, Ca), transition metals (e.g., Sc, Y), post-transition metals (e.g., Al, Ga, In), and metalloids (e.g., Ge, As, Sb). 
DFT calculations are performed to generate a dataset for formation energies in the neutral ($q=0$) and $q=+1$ charge states as well as the +1/0 charge transition level. 
We then identify key descriptors affecting formation energy and develop predictive models for the formation energies and charge transition levels of dopants in CsSnI$_3$. 
Linear and nonlinear regression models including linear regression, gaussian process regression, kernel ridge regression, and random forest regression are trained.
We also analyze the feature correlations and feature importance and extend predictions to other out-of-sample dopants in CsSnI$_3$. 

\section{\label{computational_details}Computational details}

DFT calculations were performed using the Vienna Ab initio Simulation Package (VASP).\cite{29-vasp} 
Projector-augmented wave (PAW) pseudopotentials\cite{30-pseudoProjectoraugmented-wavemethod} were employed with a plane-wave energy cutoff of 400 eV. 
The HSE06 hybrid functional\cite{31-Hybridfunctionals-based-on-screened-Coulomb-potential} was used with a mixing parameter of 0.54, and the spin-orbit coupling was also included.
The Brillouin zone for the unit cell was sampled using a $2\times2\times2$ $\Gamma$-centered k-mesh. The atomic positions were fully relaxed until the forces were less than 0.02 eV/$\mathrm{\AA}$. 
We obtained a lattice constant of $a = 8.53$ $\mathrm{\AA}$, $b = 8.81$ $ \mathrm{\AA}$, $c = 12.34$ $ \mathrm{\AA}$, and a band gap of 1.32 eV for orthorhombic CsSnI$_3$, in good agreement with experimental values.\cite{32-egcssni3, 11-usesn3-CsSnI3:semiconductorormetal?}
For defect calculations, we used a $2\times2\times1$ supercell with a $1\times1\times2$ $\Gamma$-centered k-point grid.

\begin{figure}
    \includegraphics[width=87mm]{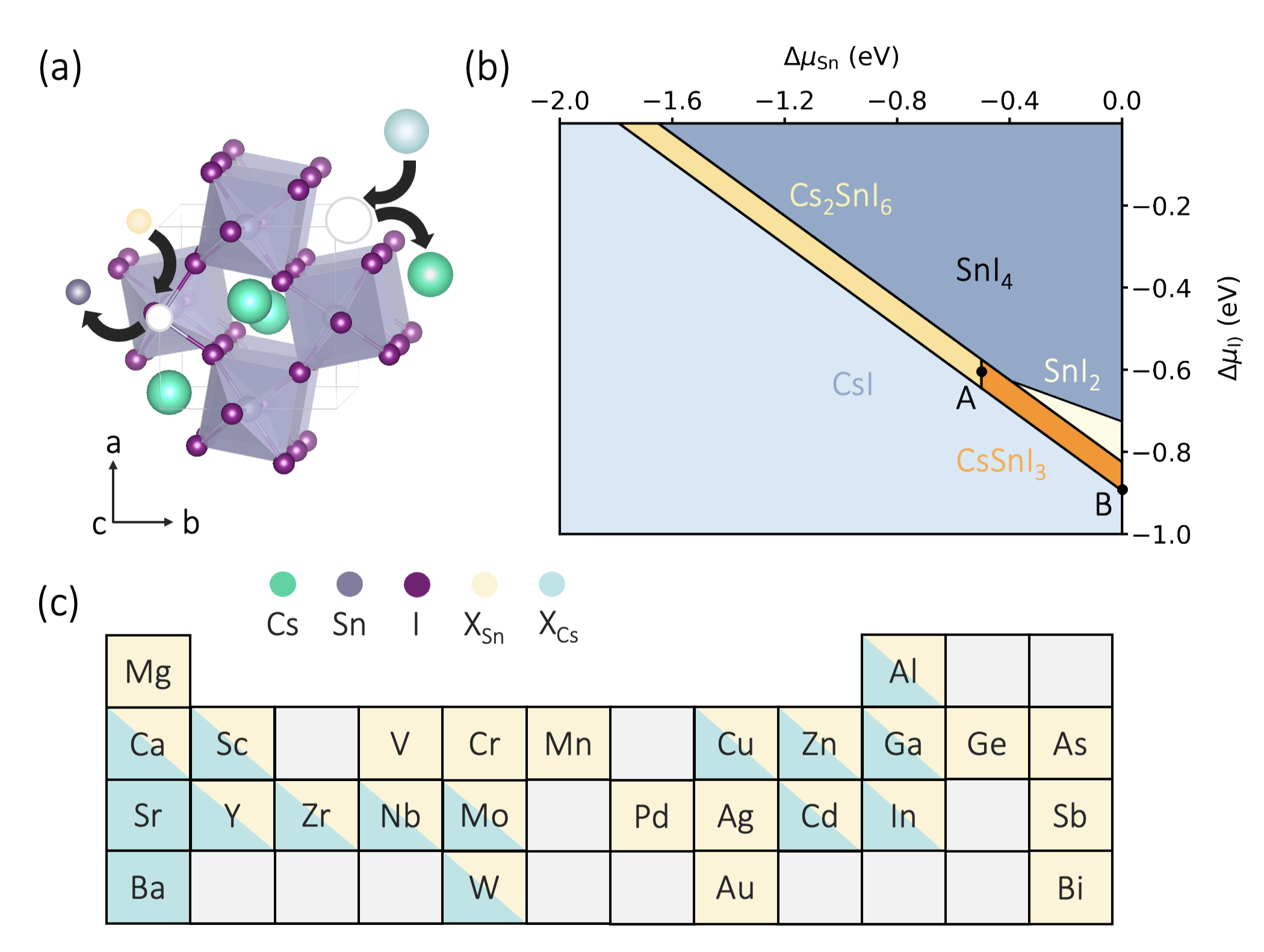}
    \caption{(a) The structure of orthorhombic CsSnI$_3$ perovskite with the Sn or Cs site substituted by dopants. 
    (b) The thermodynamically stable region for CsSnI$_3$ is shown in orange.
    Points A and B mark I-rich (Sn-poor) and I-poor (Sn-rich) conditions and the chemical potentials for Sn ($\Delta\mu_\mathrm{Sn}$) and I ($\Delta\mu_\mathrm{I}$).
    (c) The dopants calculated by DFT.}
    \label{Fig1}
\end{figure}

The formation energy of a substitutional dopant X on the Sn-site ($\mathrm{X_{Sn}}$) with the charge state of \textit{q} is calculated by
\begin{multline}
E^\mathrm{f}[\mathrm{X}_\mathrm{Sn}^q] = 
E_{\mathrm{tot}}[\mathrm{X}_\mathrm{Sn}^q] - E_{\mathrm{tot}}[{\mathrm{bulk}}] 
+ \mu_{\mathrm{Sn}}\\ - \mu_{\mathrm{X}}
+ \textit{q}(E_\mathrm{F}+E_\mathrm{vbm}) + E_\mathrm{corr}
\label{eq:formation_energy}
\end{multline}
$E_{\mathrm{tot}}[\mathrm{X}_\mathrm{Sn}^q]$ is the total energy of the supercell containing the substitutional dopant X at charge state $q$.
$E_{\mathrm{tot}}[{\mathrm{bulk}}]$ is the total energy of the perfect supercell.
$E_\mathrm{F}$ is the Fermi level and $E_\mathrm{VBM}$ is the value for the valence band maximum (VBM). 
$E_\mathrm{corr}$ is the Freysoldt’s charge correction.\cite{22-freysoldt2009fully}
$\mu_{\mathrm{Sn}}$ and $\mu_{\mathrm{X}}$ are defined as
$\mu_\mathrm{Sn}=\mu_\mathrm{Sn}^\mathrm{bulk}+\Delta\mu_\mathrm{Sn}$, and
$\mu_\mathrm{X}=\mu_\mathrm{X}^\mathrm{bulk}+\Delta\mu_\mathrm{X}$.
$\mu_\mathrm{Sn}^\mathrm{bulk}$ and $\mu_\mathrm{X}^\mathrm{bulk}$ are the single atom energy of the bulk Sn and the dopants.
$\Delta\mu_\mathrm{Sn}$ and $\Delta\mu_\mathrm{X}$ are the chemical potentials of Sn and dopants defined by the thermodynamic equilibrium condition of CsSnI$_3$, and against the formation of the competing secondary phases including CsI, SnI$_2$, SnI$_4$, and Cs$_2$SnI$_6$.
\begin{equation}
\begin{gathered} 
\Delta\mu_\mathrm{Cs}+\Delta\mu_\mathrm{Sn}+3\Delta\mu_\mathrm{I}=\Delta{H}_\mathrm{CsSnI_3} (-5.51\: \mathrm{ eV}),\\
\Delta\mu_\mathrm{Cs}+\Delta\mu_\mathrm{I}<\Delta{H}_\mathrm{CsI} (-3.72\: \mathrm{ eV}), \\
\Delta\mu_\mathrm{Sn}+2\Delta\mu_\mathrm{I}<\Delta{H}_\mathrm{SnI_2} (-1.65\: \mathrm{ eV}), \\ 
\Delta\mu_\mathrm{Sn}+4\Delta\mu_\mathrm{I}<\Delta{H}_\mathrm{SnI_4} (-2.92\: \mathrm{ eV}), \\
2\Delta\mu_\mathrm{Cs}+\Delta\mu_\mathrm{Sn}+6\Delta\mu_\mathrm{I}<\Delta{H}_\mathrm{Cs_2SnI_6} (-10.52\: \mathrm{ eV}).
\label{eq:secondary_phases}
\end{gathered}
\end{equation}
The numbers in parentheses are the calculated formation enthalpy of the secondary phases using HSE06+SOC.
This thermodynamically stable domain of CsSnI$_3$ is illustrated in orange in Figure \ref{Fig1} (b), which is consistent with previous reports\cite{35-zhang2022origins-Originsofp-Dopingincssni3}. 
The chemical potentials for I, Sn, and Cs are -0.605 eV, -0.50 eV, and -3.20 eV under  I-rich (Sn-poor) condition (point A) and -0.89 eV, 0 eV, and -2.84 eV under I-poor (Sn-rich) condition (point B).
We note that $\Delta\mu_\mathrm{X}$'s are also determined by the formation of the competing phases XI$_n$'s, where $n$ depends on the oxidation state of the dopant.
The data for the formation enthalpy of the XI$_n$'s are made available in the section Data and Code Availability.
The charge transition level (CT) from one charged state ($q_1$) to another ($q_2$) is defined as   
\begin{equation}
    CT({\textit{q}_1}/{\textit{q}_2}) = 
    \frac{E^\mathrm{f}[\mathrm{X}_\mathrm{Sn}^{q_1}, E_\mathrm{F}=0] - 
    E^\mathrm{f}[\mathrm{X}_\mathrm{Sn}^{q_2}, E_\mathrm{F}=0]}{{{\textit{q}_2}-{\textit{q}_1}}}
\end{equation}
Here, $E^\mathrm{f}[\mathrm{X}_\mathrm{Sn}^{q_1}, E_\mathrm{F}=0]$ and $E^\mathrm{f}[\mathrm{X}_\mathrm{Sn}^{q_2}, E_\mathrm{F}=0]$ are the formation energies calculated at $E_\mathrm{F} = 0$ for the defect in different charge states. 
The same approach is applied to calculate the formation energy and charge transition level of intrinsic defects in CsSnI$_3$.

\section{\label{results}Results and discussions}
\subsection{\label{dft}Defect formation energy and charge transition level}

We performed DFT calculations to obtain the formation energies ($E^\mathrm{f}$) and charge transition levels (CT) for 24 dopants substituting at the Sn site and 15 dopants substituting at the Cs site in CsSnI$_3$, aiming to identify elements that can suppress $p$-type self-doping.
The dopants we calculated are listed in Figure~\ref{Fig1} (c), including 4 alkaline earth metals, 15 transition metals, 4 post-transition metals, and 3 metalloids.
Our search for X$_\mathrm{Sn}$ is mainly focused on the trivalent dopants, which are expected to be stable at $q=+1$ under a wide range of the Fermi level, and bivalent dopants, which are expected to be stable at $q=0$ under a wide range of the Fermi level.
These dopants tend to have shallow or no charge transition levels in the band gap.\cite{18-irham2022toward,gregori2024trivalent_doping_MASnI3,gregori2024trivalent_doping_MASnIBr}
To reveal the key features that determine $E^\mathrm{f}$ and CT, we also calculated the dopants with different oxidation states, such as Zr, Nb, and Bi.

\begin{figure} 
    \includegraphics[width=87mm]{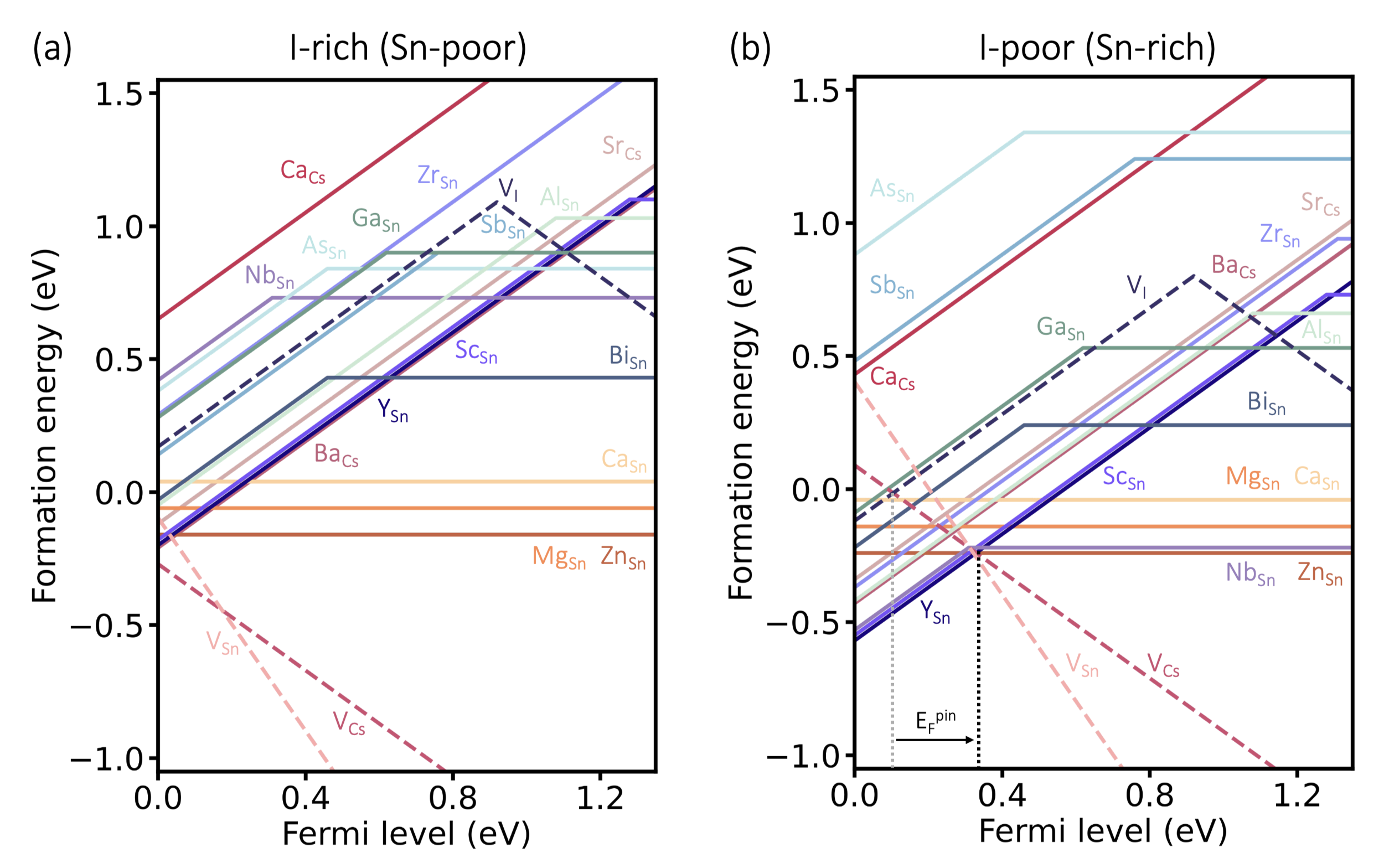} 
    \caption{The calculated defect formation energy diagrams as a function of the Fermi level for native point defects (dotted lines) and substitutional dopants (solid lines) in CsSnI$_3$ under (a) I-rich (Sn-poor) and (b) I-poor (Sn-rich) conditions. The vertical dashed lines (black and grey) indicate the pinned Fermi level (E$_\mathrm{F}^{\text{pin}}$).} 
    \label{Fig2} 
\end{figure}

Figure~\ref{Fig2} includes the intrinsic defects and the dopants with relatively low formation energies under both I-rich [Figure~\ref{Fig2} (a)] and I-poor [Figure~\ref{Fig2} (b)] conditions.
Under the I-rich condition, the $E_\mathrm{F}$ determined by the native defects in CsSnI$_3$ is pinned within the valence band (VB).
At VBM, the Cs vacancy (V$_{\text{Cs}}$) at $q=-1$ has the lowest formation energy, indicating the origin of the \textit{p}-type self-doping is primarily driven by V$_{\text{Cs}}$, consistent with previous studies.\cite{35-zhang2022origins-Originsofp-Dopingincssni3} 
Our DFT study of the CsSnI$_3$ surface phase diagram also shows that surfaces with Cs vacancies are stable under I-rich conditions.\cite{li2024cssni3_surface_dft}
Among the calculated dopants, Y$_{\text{Sn}}$ at $q=+1$ has the lowest formation energy at VBM.
However, under the I-rich condition, the formation energy of Y$_{\text{Sn}}$ at $q=+1$ is still higher than V$_{\text{Cs}}$ at $q=-1$. 
Therefore, the Fermi level cannot be shifted to a higher energy under I-rich conditions.

The I-poor condition is preferred to suppress the $p$-type self-doping.
The $E_\mathrm{F}$ determined by the native defects is pinned at 0.11 eV above the VBM under the I-poor condition: V$_{\text{Cs}}$ with $q=-1$ is compensated by the I vacancy, which prefers $q=+1$ near VBM.
We identified three trivalent elements Al, Sc, and Y that can pin the $E_\mathrm{F}$ to higher energies, which are 0.27, 0.32, and 0.33 eV above VBM.
Y$_{\text{Sn}}$ is only stable at $q=+1$ in the band gap while Al$_{\text{Sn}}$ and Sc$_{\text{Sn}}$ have a shallow CT(+1/0) near CBM.
We confirmed electron localization\cite{mu2022dopedAlGaO} at the neutral charge state (Figure S1). 
For example, Figure S1 (a) corresponds to the ground state of Al$_{\text{Sn}}$ with the charge localized near the defect while Figure S1 (b) represents a metastable state that is 0.40 eV higher in energy, where the charge is delocalized.
When Sn is substituted by bivalent elements including Mg and Zn, the defect is only stable in the neutral charge state and has relatively low formation energies.
We also identified two dopants with higher oxidation states (Zr$_{\text{Sn}}$ and Nb$_{\text{Sn}}$) that pin the Fermi level above the VBM ($\sim0.2 - 0.3$ eV). 

Zr$_{\text{Sn}}$ and Nb$_{\text{Sn}}$ are stable in the +1 charge state near the VBM, while Nb$_{\text{Sn}}$ prefers the neutral charge state across a wide range of the Fermi level. This results in a relatively deeper charge transition level (0.30 eV) within the gap compared to dopants with an oxidation state of 2 or 3. 
However, we note that deep defects may lead to slow nonradiative recombination rates due to the anharmonicity in perovskite materials.\cite{23-zhang2023iodine}

We now analyze the elemental descriptors of the substitutional dopants that correlate with the target properties including $E^\mathrm{f}$ ($q=0$), $E^\mathrm{f}$ ($q=+1$), and CT(+1/0) of X$_\mathrm{Sn}$, aiming to identify key features to predict these properties.
The oxidation state (OS) is an important feature that determines both $E^\mathrm{f}$ and CT(+1/0). 
For elements with OS=3, the formation energy at $q=0$ is higher than the bivalent elements like Zn, Mg, and Ca.
The (+1/0) charge transition levels are located near or above CBM for trivalent elements and located below VBM for bivalent elements.

For certain elements with the same OS, there is a direct trend between the atomic radius (AR) of the elements and $E^\mathrm{f}$ at both charge states. 
For example, for Zn, Mg, and Ca with OS$=$2, the $E^\mathrm{f}$ at $q=0$ under the I-rich condition increases [Zn$_{\text{Sn}}$ (-0.16 eV) $<$ Mg$_{\text{Sn}}$ (-0.06 eV) $<$ Ca$_{\text{Sn}}$ (0.04 eV)] while the atomic radius increases from Zn (1.42 \AA), Mg (1.45 \AA) to Ca (1.94 \AA).
The trend is consistent for the elements with OS greater than +2.
For example,  Al has a smaller atomic radius (1.18 \AA) than Zr (2.06 \AA) and Al has a lower formation energy than Zr in both charge states under I-rich conditions.
The Goldschmidt tolerance factor (t)\cite{goldschmidt1926gesetze} can be calculated using AR as
\begin{equation}
    t = \frac{r_{\mathrm{Cs}} + r_{\mathrm{I}}}{\sqrt{2}(r_{\mathrm{X}} + r_{\mathrm{I}})}
\end{equation}
where $r_\mathrm{Cs}$, $r_\mathrm{X}$, and $r_\mathrm{I}$ are the atomic radii of the Cs, X, and I atoms. This factor shows an inverse trend with $E^\mathrm{f}$. 
Additionally, we find that the octahedral factor ($u = r_{\mathrm{X}}$/${r_{\mathrm{I}}}$)\cite{li2008formability} calculated using Shannon’s ionic radii (IR)\cite{shannon1976IR} also shows an inverse trend with $E^\mathrm{f}$. 
Moreover, we also find that the density (D) also shows a direct trend with $E^\mathrm{f}$.
The observed trends between AR, D, t, and $u$ and the formation energies are provided in Figure S2.
We note that most of these dopants have a larger AR than Sn.

\begin{figure*} 
    \centering 
    \includegraphics[width=7in]{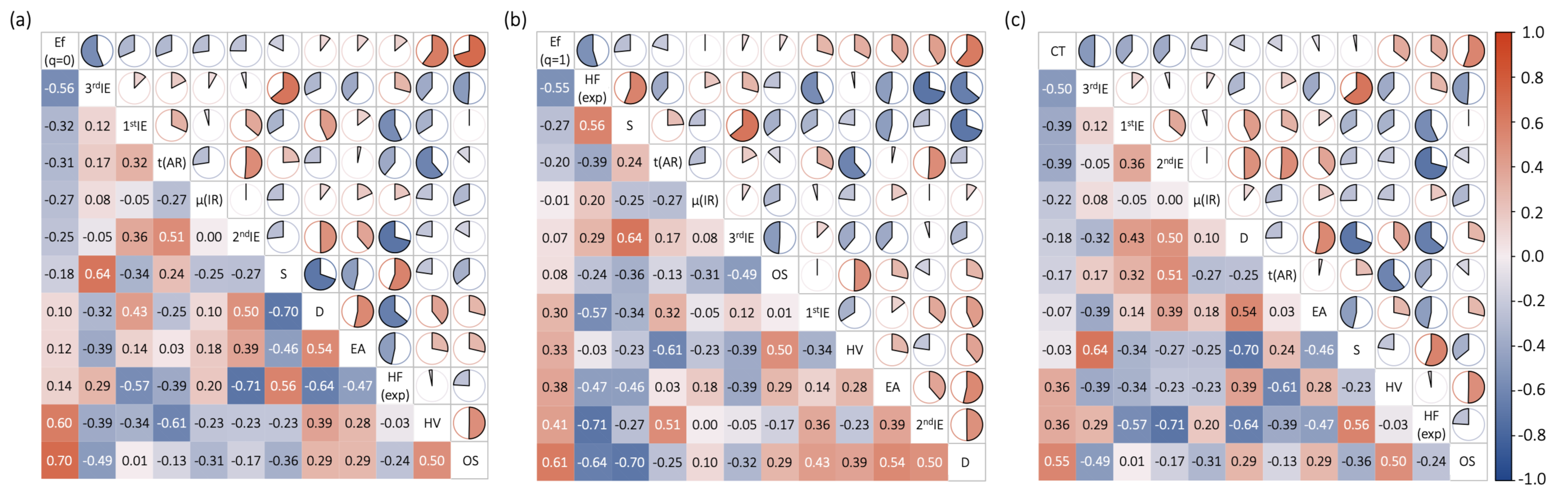} 
    \caption{Pearson correlation matrix capturing pairwise feature–feature and target–feature correlations for the X$_{\text{Sn}}$ dataset. The target properties (a) $E^f$ ($q = 0$), (b) $E^f$ ($q = +1$), and (c) CT(+1/0) along with the down-selected feature sets are listed on the diagonal. The upper and lower triangular regions of the plot convey the same information in two different visualization schemes. The filled fraction of the pie charts in the upper triangle represents the absolute value of the associated Pearson correlation coefficient, while the lighter and darker shades of color correspond to the strength of the correlation. The matrix of target property-feature correlations ranges from negative to positive correlation, from left to right, or from top to bottom.
    The features are arranged from strong negative to strong positive correlations, left to right (or top to bottom).} 
    \label{Fig3} 
\end{figure*}

Electron negativity (EN), ionization energy (IE), and electron affinity (EA) of the dopants play important roles in determining CT(+1/0).
For Ca, Mg, and Cu with OS$=+2$, CT(+1/0) of these dopants are below the VBM following the trend Cu$_{\text{Sn}}$ (-0.70 eV) $<$ Mg$_{\text{Sn}}$ (-0.30 eV) $<$ Ca$_{\text{Sn}}$ (-0.27 eV) and negatively correlated with EN of Cu (0.97) $>$ Mg (0.67) $>$ Ca (0.51).
For TMs, CT(+1/0) decreases while the first, second, and third IE increase. 
For example, CT(+1/0) of Zr$_{\text{Sn}}$, Nb$_{\text{Sn}}$, and Zn$_{\text{Sn}}$ are 1.31 eV, 0.30 eV, and -0.27 eV respectively, with 1st, 2nd, and 3rd IE increases from Zr, Nb, to Zn. 
A similar trend is observed in electron affinity (EA). For instance, the CT(+1/0) levels of Cu$_{\text{Sn}}$, Cr$_{\text{Sn}}$, and Zr$_{\text{Sn}}$ are -0.70 eV, -0.29 eV, and 1.31 eV, respectively, with EA decreasing accordingly.
The observed correlations of EN, IE, and EA with the charge transition levels are plotted in Figure S3.

In summary, we propose that trivalent dopants including Al, Sc, and Y can raise the Fermi level and suppress the $p$-type doping of CsSnI$_3$, with Y$_\mathrm{Sn}$ exhibiting the lowest formation energy under I-poor conditions. 
Dopants with higher oxidation states, such as Zr and Nb are energetically favorable at $q = +1$ near the VBM, which also raise $E_\mathrm{F}$ to higher values.
We also find that formation energies are correlated with properties including the oxidation state, tolerance factor, octahedral factor, and density. 
Charge transition levels are more correlated with elemental properties including oxidation state, electronegativity, ionization energy, and electron affinity. 
These observations will guide us in determining features for property predictions using machine learning algorithms.

\subsection{\label{features}Features for machine learning}

We initially selected 18 features representing atomic and bulk properties of the substitutional dopants and the corresponding iodide compounds (XI$_\mathrm{n}$). 
Each feature is expressed as the ratio of the dopant property to the corresponding property of Sn.
The atomic and bulk features include the ratios of electronegativity (EN), electron affinity (EA), ionization energy (IE) (including the 1st, 2nd, and 3rd IE), Pauling electronegativity (X), density (D), atomic weight (M), atomic radius (AR), covalent radius (CR), Shannon’s ionic radius (IR), and oxidation state (OS) in its most thermodynamically stable substitutional form. 
We also considered the dopant atomic features including octahedral factor ($u$), tolerance factor (t), specific heat (S), and heat of vaporization (HV), and thermodynamic properties of XI$_\mathrm{n}$ including the heat of formation (HF) from HSE06-SOC calculations [HF(cal)] and experiments [HF(exp)]. 

We used the Pearson correlation coefficient ($p$) to identify the features with strong linear correlations with properties and the highly correlated features.\cite{cohen2009pearson} 
If two features have a high absolute Pearson correlation coefficient ($|p| > 0.8$), the one with a low correlation with the property is eliminated from the feature list. 
In total, 11 features were selected for the ML model training. 
The correlations between these features and the target properties are shown in Figure \ref{Fig3}. 
The heatmaps illustrate the relationships between the down-selected features and target properties including $E^\mathrm{f}$ ($q = 0$) [Figure \ref{Fig3} (a)], $E^\mathrm{f}$ ($q = +1$) [Figure \ref{Fig3} (b)], and CT(+1/0) [Figure \ref{Fig3} (c)] under the I-rich condition. 

\begin{figure*}
    \centering
    \includegraphics[width=7in]{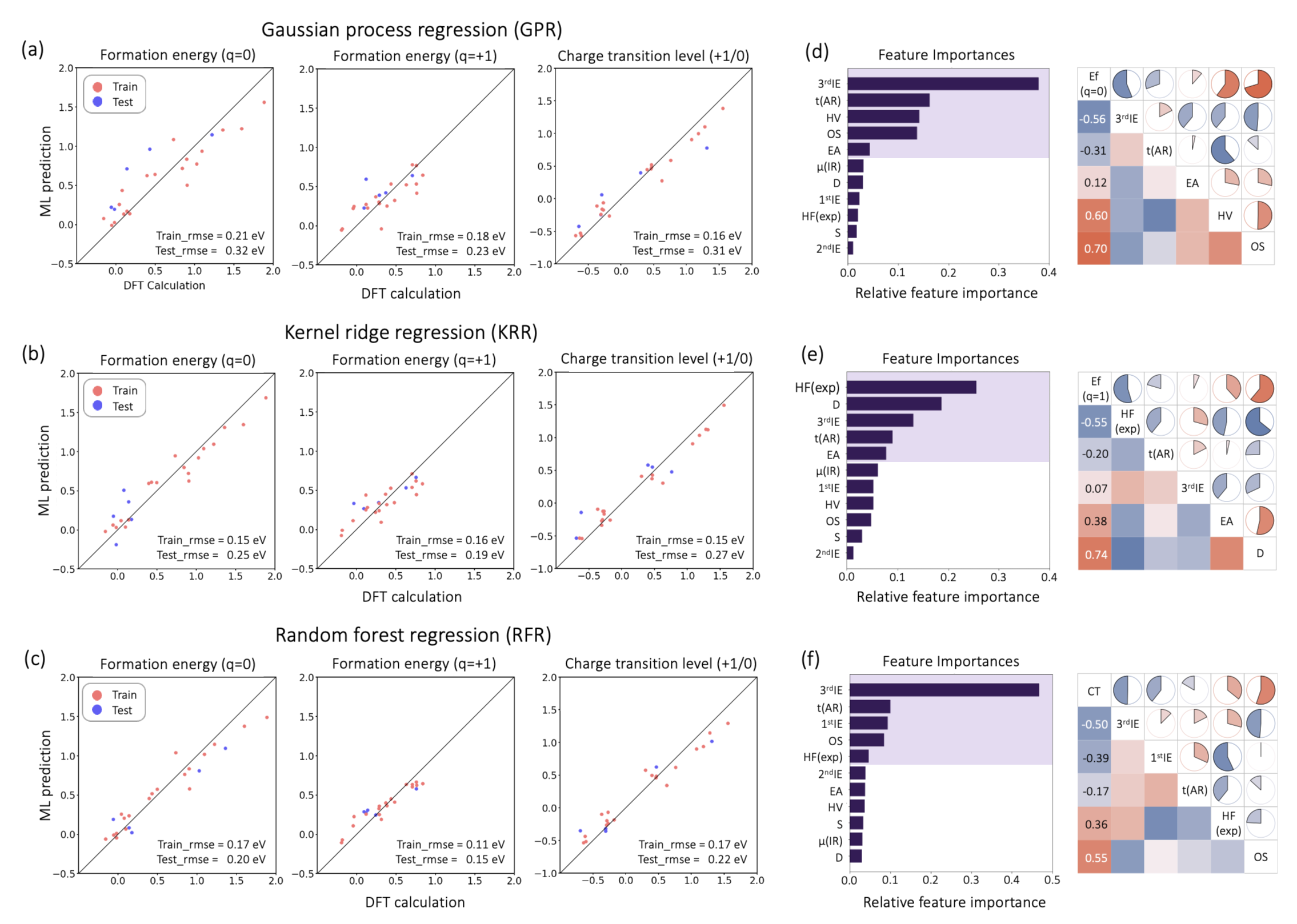}
    \caption{Parity plots from (a) Gaussian Process Regression, (b) Kernel Ridge Regression, and (c) Random Forest Regression. The relative feature importance from Random Forest Regression for predicting (d) formation energy ($q=0$), (e) formation energy ($q=+1$), and (f) charge transition level [CT(+1/0)]. The shaded purple region highlights the top five most important features.}
    \label{Fig4}
\end{figure*}

For $E^\mathrm{f}$ at $q = 0$, HV has a positive Pearson correlation coefficient ($p=0.60$) with the target property while t(AR) has a negative value of $p=-0.31$, which is consistent with our observation in Section \ref{dft}.
Additionally, the 3rd IE has a strong negative correlation ($p=-0.56$) and the OS has a strong positive correlation ($p=0.70$) with $E^\mathrm{f}$ at $q = 0$.
For $E^\mathrm{f}$ at $q = +1$, stronger correlations were observed across most features compared to the other two target properties. 
Specifically, the HF (exp) exhibited a strong negative correlation ($p=-0.55$), while D showed a strong positive correlation ($p=0.61$). 
t(AR) has a correlation of $-0.20$, which is close to the correlation observed in $E^\mathrm{f}$ at $q = 0$ ($-0.31$). 
These results indicate that structural stability and physical properties of the dopants are important descriptors to predict $E^\mathrm{f}$.

For CT(+1/0), the features with the strongest negative and positive correlations align with those observed in $E^\mathrm{f}$ at $q = 0$, as indicated by 3rd IE with $p=-0.50$ and OS with $p=0.55$
As noted in Section \ref{dft}, the EN is negatively correlated with CT(+1/0) for dopants with an oxidation state of $+2$.
In the selected feature list, EN was excluded due to its high correlation with D ($p = 0.79$) and the HF (exp) ($p = -0.87$) and its relatively small variance compared to other elemental properties.

After down-selecting the key features, we trained four machine learning (ML) algorithms including linear regression (LR), gaussian process regression (GPR), kernel ridge regression (KRR), and random forest regression (RFR) on our DFT dataset to explore their predictive capabilities. 
In our study, we used the scikit-learn package \cite{pedregosa2011scikit} to train the ML models.
We followed standard practices to split the data into training (80\%) and testing (20\%), apply grid-based hyperparameter search, and employ five-fold cross-validation to reduce overfitting.\cite{25-predictml2impuritiesinhalideperovskites, 26-predictml3predictabo3}
Model performance was evaluated using root mean square error (RMSE) as the key metric.
Additionally, we also evaluate feature importance and compare it with the Pearson correlation coefficients.

\subsection{\label{ML}Training machine learning models}

We first applied the linear regression (LR) model to predict the defect formation energies at $q=0$ and $q=+1$ and CT(+1/0). 
The parity plots, training/testing RMSE, and feature importance for the LR model are provided in Figure S4.
The RMSE values for the training/testin{g data sets were 0.23 / 0.44 eV for \(E^\mathrm{f} (q=0)\), 0.16 / 0.31 eV for \(E^\mathrm{f} (q=+1)\), and 0.31 / 0.45 eV for CT, respectively.
Compared with the nonlinear model that will be discussed later, the RMSE of the LR model is higher.
Our findings for LR align with previous studies that used linear models to predict defect properties in halide perovskites, where linear regression gives higher RMSEs as compared to nonlinear methods \cite{25-predictml2impuritiesinhalideperovskites}.
This highlights the necessity for nonlinear models to fully capture the complexity of defect features and properties.

Gaussian process regression (GPR) is known for modeling complex nonlinear correlations, employing the kernels to define a function based on the covariance of the prior distribution over the target functions.\cite{gaussian, bay}
We explored five types of kernels and the corresponding hyperparameters, alpha (the regularization parameter), and length to optimize model performance.
The kernel functions include Radial Basis Function, ExpSineSquared, Rational Quadratic, DotProduct, and Matern.\cite{duvenaud2014automatic}
Hyperparameter optimization was performed using the randomized search method.
The optimized hyperparameters are listed in Table S1. 
The parity plots using GPR are presented in Figure \ref{Fig4} (a), yielding training/testing RMSE values of 0.21 / 0.32 eV for \(E^\mathrm{f} (q=0)\), 0.18 / 0.23 eV for \(E^\mathrm{f} (q=+1)\), and 0.16 / 0.31 eV for CT.
GPR outperformed LR for all three target properties, indicating its effectiveness in capturing the underlying relationships in the data.

Kernel Ridge Regression (KRR) is also a nonlinear regression model integrating ridge regression with kernel functions.\cite{vovk2013kernel} 
The same kernel functions were tested as in GPR.
The best estimators for KRR result in RMSE values of 0.15 / 0.25 eV for \(E^\mathrm{f}(q=0)\), 0.16 / 0.19 eV for \(E^\mathrm{f}(q=+1)\), and 0.15 / 0.27 eV for CT, as shown in Figure \ref{Fig4} (b).

Random Forest Regression (RFR) is a widely used machine learning technique that combines multiple decision trees into an ensemble of predictors \cite{rfr}.
Training the RFR model involves optimizing hyperparameters including the number of trees (or estimators), maximum tree depth, number of leaf nodes, and the maximum number of features used to split a tree. 
The best hyperparameters that yielded the best predictions for all regressions are listed in Table S1.
The parity plots from the RFR model are shown in Figure \ref{Fig4} (c).
The RMSE for the training/testing datasets are 0.17 / 0.20 eV for \(E^\mathrm{f}(q=0)\), 0.11 / 0.15 eV for \(E^\mathrm{f}(q=+1)\), and 0.17 / 0.22 eV for CT, respectively.
These results demonstrate improved predictions for both \(E^f\) and CT(+1/0) compared to those achieved by using LR, GPR, and KRR.

During the training of the RFR model, we also assessed the feature importance for the three target properties [Figure \ref{Fig4} (d-f)].
For \(E^\mathrm{f}(q=0)\) [Figure \ref{Fig4} (d)], the top five most important features from the RFR training are 3rd IE, t(AR), HV, OS, and EA.
For \(E^\mathrm{f}(q=+1)\) [Figure \ref{Fig4} (e)], the top five most important features include HF(exp), D, 3rd IE, t(AR), and EA.
The feature importance for predicting formation energy for both charge states highlights three important features: 3rd IE, t(AR), and EA.
These features exhibit relatively strong positive or negative Pearson correlations in Figure \ref{Fig3} and partially overlap with the top important features predicting \(E^\mathrm{f}\) of neutral defects in ABO$_3$\cite{26-predictml3predictabo3}.
For CT(+1/0) [Figure \ref{Fig4} (f)], the top import features include 3rd IE, t(AR), 1st IE, OS, and HF(exp).
These features are also consistent with the highly correlated features shown in Figure \ref{Fig3} (c).

\subsection{\label{RFR_prediction} Prediction with random forest regression}

We also trained RFR and KRR using the formation energies of X$_{\text{Sn}}$ under the I-poor condition, aiming to directly predict out-of-sample dopants that can suppress the $p$-type self-doping.
The RMSE values for the training/testing datasets of RFR are 0.22 / 0.28 eV for \(E^\mathrm{f}(q=0)\) and 0.16 / 0.21 eV for \(E^\mathrm{f}(q=+1)\), which are lower than KRR as shown in Figure S5. 
The top important features for $q=0$ [3rd IE, HV, and OS] and $q=+1$ [HF(exp), D, and 3rd IE] remain consistent with the top important features derived from the formation energies calculated under the I-rich condition [Figure \ref{Fig4} (d-e)].
We apply the trained RFR model to predict the formation energies of 23 out-of-sample dopants under the I-poor condition.
The formation energies for \( q=0 \) and \( q=+1 \) are provided in Table S2.
Our predictions indicate that there are three trivalent dopants (La, Ce, and Pr) with the \( E^f (q=+1) \) lower than that of V$_\text{I}$ at $q=+1$ and the CT(+1/0) level is close to CBM. 
This suggests that these dopants can shift the Fermi level, pinning it closer to the conduction band compared to the intrinsic Fermi level at 0.11 eV above VBM. 
These predictions are in good agreement with previous calculations using the HSE06 functional, which show that La and Ce doping in MASnI$_3$ raises the Fermi level due to the low formation energy at q=+1 and no charge transition level in the band gap.\cite{gregori2024trivalent_doping_MASnI3}
Additionally, Sr and Ba with OS=2 have relatively low formation energies at q=0, which is consistent with previous DFT calculations using the HSE06 functional, confirming the predictive capability of the RFR model for formation energy.\cite{17-Simultaneous-suppression-of-p-doping-dft-bacs}

\subsection{\label{RFR_both_sites} Substitutional dopants on both Sn and Cs sites from random forest regression}

\begin{figure}
    \includegraphics[width=87mm]{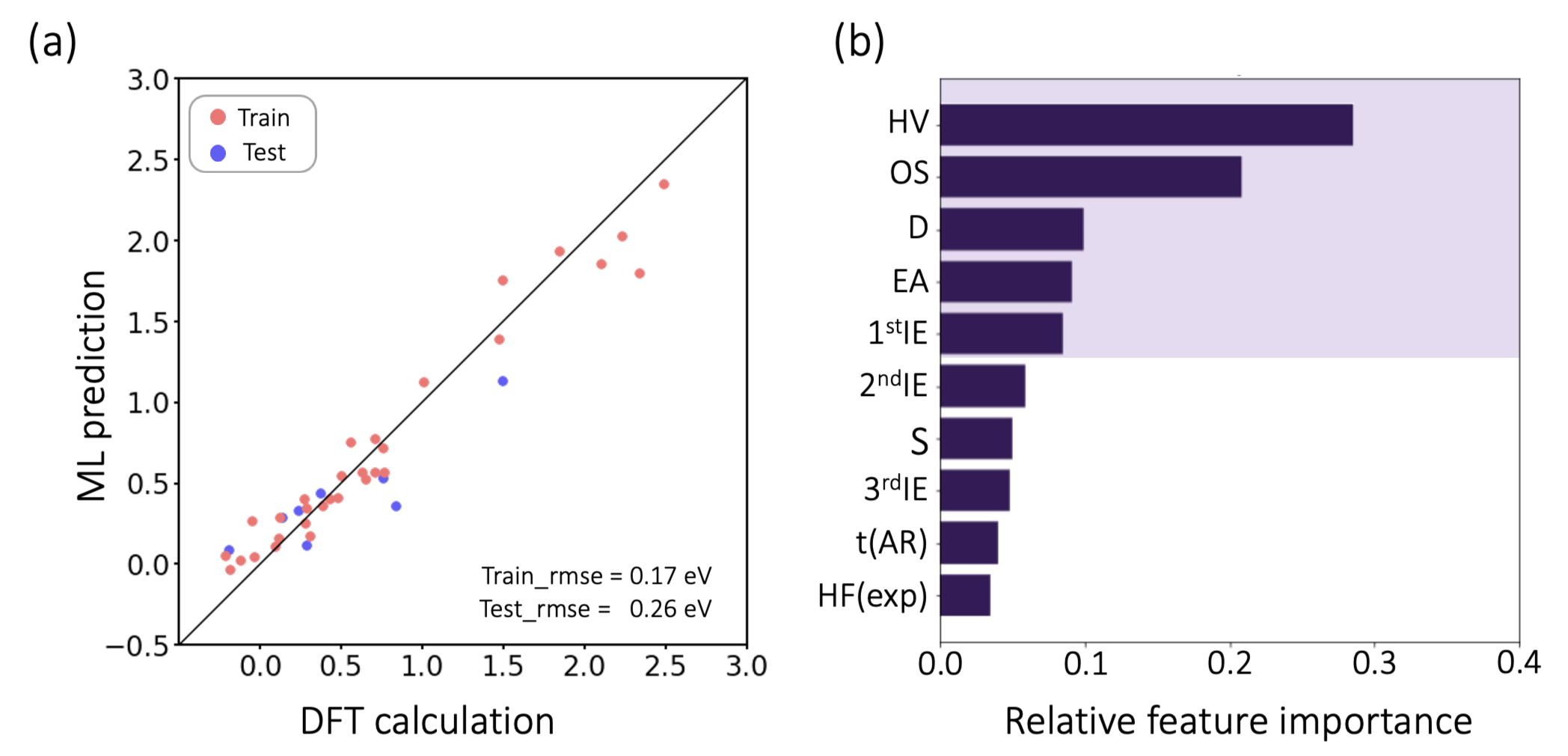}
    \caption{(a) Parity plot obtained from Random Forest Regression and (b) highlights the top five most important features for the formation energy ($q=+1$) of X$_{\text{Cs}}$ and X$_{\text{Sn}}$.}
    \label{Fig5}
\end{figure}

We performed DFT calculations of 15 substitutional dopants on the Cs site.
Ba$_\mathrm{Cs}$ and Sr$_\mathrm{Cs}$ are only stable at q=+1 within the band gap.
Ba$_{\text{Cs}}$ has the lowest formation energy and will pin the fermi level at 0.26 eV under the I-poor conditions, which is also consistent with the previous study on alkaline-earth metal doping at the Cs site.\cite{17-Simultaneous-suppression-of-p-doping-dft-bacs}
We also find Sr$_{\text{Cs}}$ to have low formation energy, pinning the Fermi level at 0.21 eV under the I-poor conditions.

We also applied RFR to predict the formation energy at $q=+1$ under I-rich conditions for substitutional dopants on both the Cs site (X$_{\text{Cs}}$) and the Sn site (X$_{\text{Sn}}$).
Figure \ref{Fig5} (a) shows the parity plot of DFT calculated versus the RFR predicted values for \( E^f \) (\( q=+1 \)) with a train/test RMSE of 0.17/0.26 eV. 
This RMSE is higher than that of the X$_{\text{Sn}}$ system as the features need to describe the interaction between dopants with two cation sites.
The top five features from the RFR training are shown in Figure \ref{Fig5} (b), including HV, OS, D, EA, and 1st IE. 
Two of these features [D and EA] are consistent with the top features from RFR training using only the X$_{\text{Sn}}$ data points [Figure~\ref{Fig4} (e)], indicating the consistency in feature correlations on both sites.

\section{\label{conclution}Conclusion}

In conclusion, we performed DFT calculations using the HSE06 functional with SOC to identify substitutional dopants in CsSnI$_3$ that suppress the $p$-type self-doping.
Trivalent dopants including Al$_\mathrm{Sn}$, Sc$_\mathrm{Sn}$, and Y$_\mathrm{Sn}$ prefer the +1 charge state and have shallow or no charge transition levels in the band gap, which pin the Fermi level at 0.27, 0.32, and 0.33 under the I-poor conditions.
Bivalent dopants including Mg and Zn are only stable in the neutral charge state and have low formation energies.
We also identified the dopants with a high oxidation state, Zr$_\mathrm{Sn}$ and Nb$_\mathrm{Sn}$, which can also raise the Fermi level under the I-poor condition.
For the substitutional dopants on the Cs site, we identified Ba$_{\text{Cs}}$ and Sr$_{\text{Cs}}$ that are only stable in the q=+1 charge state and can pin the Fermi level at 0.26 and 0.21 under the I-poor condition.

We explore machine learning regression algorithms and determine that the random forest regression can be used to develop a predictive model for the formation energy and charge transition levels of substitutional defects at the cation sites in CsSnI$_3$. 
By analyzing the feature correlation and feature importance from the random forest regression training, we identified key features including oxidation state, the heat of formation, density, and ionization energy as key descriptors that determine the defect formation energetics.
The trained model is also applied to predict out-of-sample dopants and predicts three dopants including La, Ce, and Pr that have low formation energies at the q=+1 charge state.
From a theoretical perspective, this study identifies key features that predict formation energy and charge transition levels. 
We believe that this predictive model will be valuable for investigating defects that suppress \textit{p}-type behavior in other Sn-based perovskite materials, and provide insights into the key elemental descriptors that determine the energetics in defect formation.

\section*{Data and code availability}
Datasets containing the defect formation energies and chemical potentials and the ML codes for training and prediction are available from \url{https://github.com/Mengen-W/Doping_CsSnI3_DFT_ML}.

\section*{Acknowledgments}
The work was supported by the new faculty start-up and Transdisciplinary Areas of Excellence (TAE) Seed Grant funds from SUNY Binghamton.
This work used Bridges-2 at Pittsburgh Supercomputing Center through allocation MAT230043 from the Advanced Cyberinfrastructure Coordination Ecosystem: Services \text{\&} Support (ACCESS) program, which is supported by National Science Foundation grants \text{\#}2138259, \text{\#}2138286, \text{\#}2138307, \text{\#}2137603, and \text{\#}2138296.
This work also used computational resources provided by the SPIEDIE cluster at the State University of New York at Binghamton.

\bibliography{references}

\providecommand{\noopsort}[1]{}\providecommand{\singleletter}[1]{#1}%
\begin{thebibliography}{46}%
\makeatletter
\providecommand \@ifxundefined [1]{%
 \@ifx{#1\undefined}
}%
\providecommand \@ifnum [1]{%
 \ifnum #1\expandafter \@firstoftwo
 \else \expandafter \@secondoftwo
 \fi
}%
\providecommand \@ifx [1]{%
 \ifx #1\expandafter \@firstoftwo
 \else \expandafter \@secondoftwo
 \fi
}%
\providecommand \natexlab [1]{#1}%
\providecommand \enquote  [1]{``#1''}%
\providecommand \bibnamefont  [1]{#1}%
\providecommand \bibfnamefont [1]{#1}%
\providecommand \citenamefont [1]{#1}%
\providecommand \href@noop [0]{\@secondoftwo}%
\providecommand \href [0]{\begingroup \@sanitize@url \@href}%
\providecommand \@href[1]{\@@startlink{#1}\@@href}%
\providecommand \@@href[1]{\endgroup#1\@@endlink}%
\providecommand \@sanitize@url [0]{\catcode `\\12\catcode `\$12\catcode `\&12\catcode `\#12\catcode `\^12\catcode `\_12\catcode `\%12\relax}%
\providecommand \@@startlink[1]{}%
\providecommand \@@endlink[0]{}%
\providecommand \url  [0]{\begingroup\@sanitize@url \@url }%
\providecommand \@url [1]{\endgroup\@href {#1}{\urlprefix }}%
\providecommand \urlprefix  [0]{URL }%
\providecommand \Eprint [0]{\href }%
\providecommand \doibase [0]{http://dx.doi.org/}%
\providecommand \selectlanguage [0]{\@gobble}%
\providecommand \bibinfo  [0]{\@secondoftwo}%
\providecommand \bibfield  [0]{\@secondoftwo}%
\providecommand \translation [1]{[#1]}%
\providecommand \BibitemOpen [0]{}%
\providecommand \bibitemStop [0]{}%
\providecommand \bibitemNoStop [0]{.\EOS\space}%
\providecommand \EOS [0]{\spacefactor3000\relax}%
\providecommand \BibitemShut  [1]{\csname bibitem#1\endcsname}%
\let\auto@bib@innerbib\@empty
\bibitem [{\citenamefont {Zhou}\ and\ \citenamefont {Zhao}(2019)}]{1-Chemica-stability}%
  \BibitemOpen
  \bibfield  {author} {\bibinfo {author} {\bibfnamefont {Y.}~\bibnamefont {Zhou}}\ and\ \bibinfo {author} {\bibfnamefont {Y.}~\bibnamefont {Zhao}},\ }\href {https://pubs.rsc.org/en/content/articlehtml/2010/g1/c8ee03559h} {\bibfield  {journal} {\bibinfo  {journal} {Energy \& Environmental Science}\ }\textbf {\bibinfo {volume} {12}},\ \bibinfo {pages} {1495} (\bibinfo {year} {2019})}\BibitemShut {NoStop}%
\bibitem [{\citenamefont {Dong}\ \emph {et~al.}(2015)\citenamefont {Dong}, \citenamefont {Fang}, \citenamefont {Shao}, \citenamefont {Mulligan}, \citenamefont {Qiu}, \citenamefont {Cao},\ and\ \citenamefont {Huang}}]{2-Electron-hole-diffusion-lengths-solution-grown-CH3NH3PbI3-prop1}%
  \BibitemOpen
  \bibfield  {author} {\bibinfo {author} {\bibfnamefont {Q.}~\bibnamefont {Dong}}, \bibinfo {author} {\bibfnamefont {Y.}~\bibnamefont {Fang}}, \bibinfo {author} {\bibfnamefont {Y.}~\bibnamefont {Shao}}, \bibinfo {author} {\bibfnamefont {P.}~\bibnamefont {Mulligan}}, \bibinfo {author} {\bibfnamefont {J.}~\bibnamefont {Qiu}}, \bibinfo {author} {\bibfnamefont {L.}~\bibnamefont {Cao}}, \ and\ \bibinfo {author} {\bibfnamefont {J.}~\bibnamefont {Huang}},\ }\href {https://www.science.org/doi/abs/10.1126/science.aaa5760} {\bibfield  {journal} {\bibinfo  {journal} {Science}\ }\textbf {\bibinfo {volume} {347}},\ \bibinfo {pages} {967} (\bibinfo {year} {2015})}\BibitemShut {NoStop}%
\bibitem [{\citenamefont {Stranks}\ \emph {et~al.}(2013)\citenamefont {Stranks}, \citenamefont {Eperon}, \citenamefont {Grancini}, \citenamefont {Menelaou}, \citenamefont {Alcocer}, \citenamefont {Leijtens}, \citenamefont {Herz}, \citenamefont {Petrozza},\ and\ \citenamefont {Snaith}}]{3-Electron-hole-diffusion-lengths-exceeding-prop2}%
  \BibitemOpen
  \bibfield  {author} {\bibinfo {author} {\bibfnamefont {S.~D.}\ \bibnamefont {Stranks}}, \bibinfo {author} {\bibfnamefont {G.~E.}\ \bibnamefont {Eperon}}, \bibinfo {author} {\bibfnamefont {G.}~\bibnamefont {Grancini}}, \bibinfo {author} {\bibfnamefont {C.}~\bibnamefont {Menelaou}}, \bibinfo {author} {\bibfnamefont {M.~J.}\ \bibnamefont {Alcocer}}, \bibinfo {author} {\bibfnamefont {T.}~\bibnamefont {Leijtens}}, \bibinfo {author} {\bibfnamefont {L.~M.}\ \bibnamefont {Herz}}, \bibinfo {author} {\bibfnamefont {A.}~\bibnamefont {Petrozza}}, \ and\ \bibinfo {author} {\bibfnamefont {H.~J.}\ \bibnamefont {Snaith}},\ }\href {https://www.science.org/doi/10.1126/science.1243982} {\bibfield  {journal} {\bibinfo  {journal} {Science}\ }\textbf {\bibinfo {volume} {342}},\ \bibinfo {pages} {341} (\bibinfo {year} {2013})}\BibitemShut {NoStop}%
\bibitem [{\citenamefont {Noh}\ \emph {et~al.}(2013)\citenamefont {Noh}, \citenamefont {Im}, \citenamefont {Heo}, \citenamefont {Mandal},\ and\ \citenamefont {Seok}}]{4-Chemical-management-for-colorful-efficient-and-stable-inorganic-organic-hybrid-nanostructured-solar-cells-prop3}%
  \BibitemOpen
  \bibfield  {author} {\bibinfo {author} {\bibfnamefont {J.~H.}\ \bibnamefont {Noh}}, \bibinfo {author} {\bibfnamefont {S.~H.}\ \bibnamefont {Im}}, \bibinfo {author} {\bibfnamefont {J.~H.}\ \bibnamefont {Heo}}, \bibinfo {author} {\bibfnamefont {T.~N.}\ \bibnamefont {Mandal}}, \ and\ \bibinfo {author} {\bibfnamefont {S.~I.}\ \bibnamefont {Seok}},\ }\href {https://pubs.acs.org/doi/10.1021/nl400349b} {\bibfield  {journal} {\bibinfo  {journal} {Nano Letters}\ }\textbf {\bibinfo {volume} {13}},\ \bibinfo {pages} {1764} (\bibinfo {year} {2013})}\BibitemShut {NoStop}%
\bibitem [{\citenamefont {Kojima}\ \emph {et~al.}(2009)\citenamefont {Kojima}, \citenamefont {Teshima}, \citenamefont {Shirai},\ and\ \citenamefont {Miyasaka}}]{5-pce3.8}%
  \BibitemOpen
  \bibfield  {author} {\bibinfo {author} {\bibfnamefont {A.}~\bibnamefont {Kojima}}, \bibinfo {author} {\bibfnamefont {K.}~\bibnamefont {Teshima}}, \bibinfo {author} {\bibfnamefont {Y.}~\bibnamefont {Shirai}}, \ and\ \bibinfo {author} {\bibfnamefont {T.}~\bibnamefont {Miyasaka}},\ }\href {https://pubs.acs.org/doi/full/10.1021/ja809598r} {\bibfield  {journal} {\bibinfo  {journal} {Journal of the American Chemical Society}\ }\textbf {\bibinfo {volume} {131}},\ \bibinfo {pages} {6050} (\bibinfo {year} {2009})}\BibitemShut {NoStop}%
\bibitem [{6-p()}]{6-pce25}%
  \BibitemOpen
  \href@noop {} {\enquote {\bibinfo {title} {National renewable energy laboratory (nrel). best research-cell efficiency chart. (2022).}}\ }\bibinfo {howpublished} {\url{https://www.nrel.gov/pv/cell-efficiency.html}}\BibitemShut {NoStop}%
\bibitem [{\citenamefont {Chung}\ \emph {et~al.}(2012{\natexlab{a}})\citenamefont {Chung}, \citenamefont {Lee}, \citenamefont {He}, \citenamefont {Chang},\ and\ \citenamefont {Kanatzidis}}]{10-usesn2-All-solid-state}%
  \BibitemOpen
  \bibfield  {author} {\bibinfo {author} {\bibfnamefont {I.}~\bibnamefont {Chung}}, \bibinfo {author} {\bibfnamefont {B.}~\bibnamefont {Lee}}, \bibinfo {author} {\bibfnamefont {J.}~\bibnamefont {He}}, \bibinfo {author} {\bibfnamefont {R.~P.}\ \bibnamefont {Chang}}, \ and\ \bibinfo {author} {\bibfnamefont {M.~G.}\ \bibnamefont {Kanatzidis}},\ }\href {https://www.nature.com/articles/nature11067} {\bibfield  {journal} {\bibinfo  {journal} {Nature}\ }\textbf {\bibinfo {volume} {485}},\ \bibinfo {pages} {486} (\bibinfo {year} {2012}{\natexlab{a}})}\BibitemShut {NoStop}%
\bibitem [{\citenamefont {Chung}\ \emph {et~al.}(2012{\natexlab{b}})\citenamefont {Chung}, \citenamefont {Song}, \citenamefont {Im}, \citenamefont {Androulakis}, \citenamefont {Malliakas}, \citenamefont {Li}, \citenamefont {Freeman}, \citenamefont {Kenney},\ and\ \citenamefont {Kanatzidis}}]{11-usesn3-CsSnI3:semiconductorormetal?}%
  \BibitemOpen
  \bibfield  {author} {\bibinfo {author} {\bibfnamefont {I.}~\bibnamefont {Chung}}, \bibinfo {author} {\bibfnamefont {J.-H.}\ \bibnamefont {Song}}, \bibinfo {author} {\bibfnamefont {J.}~\bibnamefont {Im}}, \bibinfo {author} {\bibfnamefont {J.}~\bibnamefont {Androulakis}}, \bibinfo {author} {\bibfnamefont {C.~D.}\ \bibnamefont {Malliakas}}, \bibinfo {author} {\bibfnamefont {H.}~\bibnamefont {Li}}, \bibinfo {author} {\bibfnamefont {A.~J.}\ \bibnamefont {Freeman}}, \bibinfo {author} {\bibfnamefont {J.~T.}\ \bibnamefont {Kenney}}, \ and\ \bibinfo {author} {\bibfnamefont {M.~G.}\ \bibnamefont {Kanatzidis}},\ }\href {https://pubs.acs.org/doi/full/10.1021/ja301539s} {\bibfield  {journal} {\bibinfo  {journal} {Journal of the American Chemical Society}\ }\textbf {\bibinfo {volume} {134}},\ \bibinfo {pages} {8579} (\bibinfo {year} {2012}{\natexlab{b}})}\BibitemShut {NoStop}%
\bibitem [{\citenamefont {Eperon}\ \emph {et~al.}(2015)\citenamefont {Eperon}, \citenamefont {Habisreutinger}, \citenamefont {Leijtens}, \citenamefont {Bruijnaers}, \citenamefont {van Franeker}, \citenamefont {DeQuilettes}, \citenamefont {Pathak}, \citenamefont {Sutton}, \citenamefont {Grancini}, \citenamefont {Ginger} \emph {et~al.}}]{7-toxicpb1-The-importance-of-moisture-in-hybrid-lead-halide-perovskite}%
  \BibitemOpen
  \bibfield  {author} {\bibinfo {author} {\bibfnamefont {G.~E.}\ \bibnamefont {Eperon}}, \bibinfo {author} {\bibfnamefont {S.~N.}\ \bibnamefont {Habisreutinger}}, \bibinfo {author} {\bibfnamefont {T.}~\bibnamefont {Leijtens}}, \bibinfo {author} {\bibfnamefont {B.~J.}\ \bibnamefont {Bruijnaers}}, \bibinfo {author} {\bibfnamefont {J.~J.}\ \bibnamefont {van Franeker}}, \bibinfo {author} {\bibfnamefont {D.~W.}\ \bibnamefont {DeQuilettes}}, \bibinfo {author} {\bibfnamefont {S.}~\bibnamefont {Pathak}}, \bibinfo {author} {\bibfnamefont {R.~J.}\ \bibnamefont {Sutton}}, \bibinfo {author} {\bibfnamefont {G.}~\bibnamefont {Grancini}}, \bibinfo {author} {\bibfnamefont {D.~S.}\ \bibnamefont {Ginger}},  \emph {et~al.},\ }\href {https://pubs.acs.org/doi/full/10.1021/acsnano.5b03626} {\bibfield  {journal} {\bibinfo  {journal} {ACS Nano}\ }\textbf {\bibinfo {volume} {9}},\ \bibinfo {pages} {9380} (\bibinfo {year} {2015})}\BibitemShut {NoStop}%
\bibitem [{\citenamefont {Babayigit}\ \emph {et~al.}(2016)\citenamefont {Babayigit}, \citenamefont {Ethirajan}, \citenamefont {Muller},\ and\ \citenamefont {Conings}}]{8-toxicpb2-Toxicity-of-organometal-halide-perovskite}%
  \BibitemOpen
  \bibfield  {author} {\bibinfo {author} {\bibfnamefont {A.}~\bibnamefont {Babayigit}}, \bibinfo {author} {\bibfnamefont {A.}~\bibnamefont {Ethirajan}}, \bibinfo {author} {\bibfnamefont {M.}~\bibnamefont {Muller}}, \ and\ \bibinfo {author} {\bibfnamefont {B.}~\bibnamefont {Conings}},\ }\href {https://www.nature.com/articles/nmat4572} {\bibfield  {journal} {\bibinfo  {journal} {Nature Materials}\ }\textbf {\bibinfo {volume} {15}},\ \bibinfo {pages} {247} (\bibinfo {year} {2016})}\BibitemShut {NoStop}%
\bibitem [{\citenamefont {Yu}\ \emph {et~al.}(2021)\citenamefont {Yu}, \citenamefont {Chen}, \citenamefont {Zhu}, \citenamefont {Wang}, \citenamefont {Han}, \citenamefont {Chen}, \citenamefont {Zhang}, \citenamefont {Du},\ and\ \citenamefont {He}}]{12-pcesn}%
  \BibitemOpen
  \bibfield  {author} {\bibinfo {author} {\bibfnamefont {B.-B.}\ \bibnamefont {Yu}}, \bibinfo {author} {\bibfnamefont {Z.}~\bibnamefont {Chen}}, \bibinfo {author} {\bibfnamefont {Y.}~\bibnamefont {Zhu}}, \bibinfo {author} {\bibfnamefont {Y.}~\bibnamefont {Wang}}, \bibinfo {author} {\bibfnamefont {B.}~\bibnamefont {Han}}, \bibinfo {author} {\bibfnamefont {G.}~\bibnamefont {Chen}}, \bibinfo {author} {\bibfnamefont {X.}~\bibnamefont {Zhang}}, \bibinfo {author} {\bibfnamefont {Z.}~\bibnamefont {Du}}, \ and\ \bibinfo {author} {\bibfnamefont {Z.}~\bibnamefont {He}},\ }\href {https://onlinelibrary.wiley.com/doi/full/10.1002/adma.202102055} {\bibfield  {journal} {\bibinfo  {journal} {Advanced Materials}\ }\textbf {\bibinfo {volume} {33}},\ \bibinfo {pages} {2102055} (\bibinfo {year} {2021})}\BibitemShut {NoStop}%
\bibitem [{\citenamefont {Song}\ \emph {et~al.}(2017)\citenamefont {Song}, \citenamefont {Yokoyama}, \citenamefont {Stoumpos}, \citenamefont {Logsdon}, \citenamefont {Cao}, \citenamefont {Wasielewski}, \citenamefont {Aramaki},\ and\ \citenamefont {Kanatzidis}}]{13-nonrad-Importance-of-reducing-vapor-atmosphere}%
  \BibitemOpen
  \bibfield  {author} {\bibinfo {author} {\bibfnamefont {T.-B.}\ \bibnamefont {Song}}, \bibinfo {author} {\bibfnamefont {T.}~\bibnamefont {Yokoyama}}, \bibinfo {author} {\bibfnamefont {C.~C.}\ \bibnamefont {Stoumpos}}, \bibinfo {author} {\bibfnamefont {J.}~\bibnamefont {Logsdon}}, \bibinfo {author} {\bibfnamefont {D.~H.}\ \bibnamefont {Cao}}, \bibinfo {author} {\bibfnamefont {M.~R.}\ \bibnamefont {Wasielewski}}, \bibinfo {author} {\bibfnamefont {S.}~\bibnamefont {Aramaki}}, \ and\ \bibinfo {author} {\bibfnamefont {M.~G.}\ \bibnamefont {Kanatzidis}},\ }\href {https://pubs.acs.org/doi/full/10.1021/jacs.6b10734} {\bibfield  {journal} {\bibinfo  {journal} {Journal of the American Chemical Society}\ }\textbf {\bibinfo {volume} {139}},\ \bibinfo {pages} {836} (\bibinfo {year} {2017})}\BibitemShut {NoStop}%
\bibitem [{\citenamefont {Xu}\ \emph {et~al.}(2014)\citenamefont {Xu}, \citenamefont {Chen}, \citenamefont {Xiang}, \citenamefont {Gong},\ and\ \citenamefont {Wei}}]{14-ptype2-Influence-of-defects-and-synthesis}%
  \BibitemOpen
  \bibfield  {author} {\bibinfo {author} {\bibfnamefont {P.}~\bibnamefont {Xu}}, \bibinfo {author} {\bibfnamefont {S.}~\bibnamefont {Chen}}, \bibinfo {author} {\bibfnamefont {H.-J.}\ \bibnamefont {Xiang}}, \bibinfo {author} {\bibfnamefont {X.-G.}\ \bibnamefont {Gong}}, \ and\ \bibinfo {author} {\bibfnamefont {S.-H.}\ \bibnamefont {Wei}},\ }\href {https://pubs.acs.org/doi/full/10.1021/cm503122j} {\bibfield  {journal} {\bibinfo  {journal} {Chemistry of Materials}\ }\textbf {\bibinfo {volume} {26}},\ \bibinfo {pages} {6068} (\bibinfo {year} {2014})}\BibitemShut {NoStop}%
\bibitem [{\citenamefont {Shi}\ \emph {et~al.}(2017)\citenamefont {Shi}, \citenamefont {Zhang}, \citenamefont {Meng}, \citenamefont {Teng}, \citenamefont {Liu}, \citenamefont {Yang}, \citenamefont {Yan}, \citenamefont {Yip},\ and\ \citenamefont {Zhao}}]{15-ptype3-Effects-of-organic-cations}%
  \BibitemOpen
  \bibfield  {author} {\bibinfo {author} {\bibfnamefont {T.}~\bibnamefont {Shi}}, \bibinfo {author} {\bibfnamefont {H.-S.}\ \bibnamefont {Zhang}}, \bibinfo {author} {\bibfnamefont {W.}~\bibnamefont {Meng}}, \bibinfo {author} {\bibfnamefont {Q.}~\bibnamefont {Teng}}, \bibinfo {author} {\bibfnamefont {M.}~\bibnamefont {Liu}}, \bibinfo {author} {\bibfnamefont {X.}~\bibnamefont {Yang}}, \bibinfo {author} {\bibfnamefont {Y.}~\bibnamefont {Yan}}, \bibinfo {author} {\bibfnamefont {H.-L.}\ \bibnamefont {Yip}}, \ and\ \bibinfo {author} {\bibfnamefont {Y.-J.}\ \bibnamefont {Zhao}},\ }\href {https://pubs.rsc.org/en/content/articlelanding/2017/ta/c7ta02662e} {\bibfield  {journal} {\bibinfo  {journal} {Journal of Materials Chemistry A}\ }\textbf {\bibinfo {volume} {5}},\ \bibinfo {pages} {15124} (\bibinfo {year} {2017})}\BibitemShut {NoStop}%
\bibitem [{\citenamefont {Yu}\ \emph {et~al.}(2022)\citenamefont {Yu}, \citenamefont {Chen}, \citenamefont {Harvey}, \citenamefont {Ni}, \citenamefont {Chen}, \citenamefont {Chen}, \citenamefont {Yao}, \citenamefont {Xiao}, \citenamefont {Xu}, \citenamefont {Yang} \emph {et~al.}}]{16-yu2022gradient-Gradient-Doping-in-Sn--Pb-Perovskites}%
  \BibitemOpen
  \bibfield  {author} {\bibinfo {author} {\bibfnamefont {Z.}~\bibnamefont {Yu}}, \bibinfo {author} {\bibfnamefont {X.}~\bibnamefont {Chen}}, \bibinfo {author} {\bibfnamefont {S.~P.}\ \bibnamefont {Harvey}}, \bibinfo {author} {\bibfnamefont {Z.}~\bibnamefont {Ni}}, \bibinfo {author} {\bibfnamefont {B.}~\bibnamefont {Chen}}, \bibinfo {author} {\bibfnamefont {S.}~\bibnamefont {Chen}}, \bibinfo {author} {\bibfnamefont {C.}~\bibnamefont {Yao}}, \bibinfo {author} {\bibfnamefont {X.}~\bibnamefont {Xiao}}, \bibinfo {author} {\bibfnamefont {S.}~\bibnamefont {Xu}}, \bibinfo {author} {\bibfnamefont {G.}~\bibnamefont {Yang}},  \emph {et~al.},\ }\href {https://onlinelibrary.wiley.com/doi/full/10.1002/adma.202110351} {\bibfield  {journal} {\bibinfo  {journal} {Advanced Materials}\ }\textbf {\bibinfo {volume} {34}},\ \bibinfo {pages} {2110351} (\bibinfo {year} {2022})}\BibitemShut {NoStop}%
\bibitem [{\citenamefont {Zhang}\ and\ \citenamefont {Chen}(2023)}]{17-Simultaneous-suppression-of-p-doping-dft-bacs}%
  \BibitemOpen
  \bibfield  {author} {\bibinfo {author} {\bibfnamefont {J.}~\bibnamefont {Zhang}}\ and\ \bibinfo {author} {\bibfnamefont {L.}~\bibnamefont {Chen}},\ }\href {https://pubs.acs.org/doi/full/10.1021/acs.jpclett.3c00831} {\bibfield  {journal} {\bibinfo  {journal} {The Journal of Physical Chemistry Letters}\ }\textbf {\bibinfo {volume} {14}},\ \bibinfo {pages} {4058} (\bibinfo {year} {2023})}\BibitemShut {NoStop}%
\bibitem [{\citenamefont {Gregori}\ \emph {et~al.}(2024{\natexlab{a}})\citenamefont {Gregori}, \citenamefont {Frasca}, \citenamefont {Meggiolaro}, \citenamefont {Belanzoni}, \citenamefont {Ashraf}, \citenamefont {Musiienko}, \citenamefont {Abate},\ and\ \citenamefont {De~Angelis}}]{gregori2024trivalent_doping_MASnI3}%
  \BibitemOpen
  \bibfield  {author} {\bibinfo {author} {\bibfnamefont {L.}~\bibnamefont {Gregori}}, \bibinfo {author} {\bibfnamefont {C.}~\bibnamefont {Frasca}}, \bibinfo {author} {\bibfnamefont {D.}~\bibnamefont {Meggiolaro}}, \bibinfo {author} {\bibfnamefont {P.}~\bibnamefont {Belanzoni}}, \bibinfo {author} {\bibfnamefont {M.~W.}\ \bibnamefont {Ashraf}}, \bibinfo {author} {\bibfnamefont {A.}~\bibnamefont {Musiienko}}, \bibinfo {author} {\bibfnamefont {A.}~\bibnamefont {Abate}}, \ and\ \bibinfo {author} {\bibfnamefont {F.}~\bibnamefont {De~Angelis}},\ }\href {https://pubs.acs.org/doi/10.1021/acsenergylett.4c00840} {\bibfield  {journal} {\bibinfo  {journal} {ACS Energy Letters}\ }\textbf {\bibinfo {volume} {9}},\ \bibinfo {pages} {3036} (\bibinfo {year} {2024}{\natexlab{a}})}\BibitemShut {NoStop}%
\bibitem [{\citenamefont {Gregori}\ \emph {et~al.}(2024{\natexlab{b}})\citenamefont {Gregori}, \citenamefont {Meggiolaro},\ and\ \citenamefont {De~Angelis}}]{gregori2024trivalent_doping_MASnIBr}%
  \BibitemOpen
  \bibfield  {author} {\bibinfo {author} {\bibfnamefont {L.}~\bibnamefont {Gregori}}, \bibinfo {author} {\bibfnamefont {D.}~\bibnamefont {Meggiolaro}}, \ and\ \bibinfo {author} {\bibfnamefont {F.}~\bibnamefont {De~Angelis}},\ }\href {https://onlinelibrary.wiley.com/doi/10.1002/smll.202403413} {\bibfield  {journal} {\bibinfo  {journal} {Small}\ ,\ \bibinfo {pages} {2403413}} (\bibinfo {year} {2024}{\natexlab{b}})}\BibitemShut {NoStop}%
\bibitem [{\citenamefont {Kang}\ \emph {et~al.}(2021)\citenamefont {Kang}, \citenamefont {Li},\ and\ \citenamefont {Wei}}]{kang2021review_dft_perovskite}%
  \BibitemOpen
  \bibfield  {author} {\bibinfo {author} {\bibfnamefont {J.}~\bibnamefont {Kang}}, \bibinfo {author} {\bibfnamefont {J.}~\bibnamefont {Li}}, \ and\ \bibinfo {author} {\bibfnamefont {S.-H.}\ \bibnamefont {Wei}},\ }\href {https://pubs.aip.org/aip/apr/article/8/3/031302/124874} {\bibfield  {journal} {\bibinfo  {journal} {Applied Physics Reviews}\ }\textbf {\bibinfo {volume} {8}} (\bibinfo {year} {2021})}\BibitemShut {NoStop}%
\bibitem [{\citenamefont {Mu}\ \emph {et~al.}(2022)\citenamefont {Mu}, \citenamefont {Wang}, \citenamefont {Varley}, \citenamefont {Lyons}, \citenamefont {Wickramaratne},\ and\ \citenamefont {Van~de Walle}}]{mu2022dopedAlGaO}%
  \BibitemOpen
  \bibfield  {author} {\bibinfo {author} {\bibfnamefont {S.}~\bibnamefont {Mu}}, \bibinfo {author} {\bibfnamefont {M.}~\bibnamefont {Wang}}, \bibinfo {author} {\bibfnamefont {J.~B.}\ \bibnamefont {Varley}}, \bibinfo {author} {\bibfnamefont {J.~L.}\ \bibnamefont {Lyons}}, \bibinfo {author} {\bibfnamefont {D.}~\bibnamefont {Wickramaratne}}, \ and\ \bibinfo {author} {\bibfnamefont {C.~G.}\ \bibnamefont {Van~de Walle}},\ }\href {https://journals.aps.org/prb/abstract/10.1103/PhysRevB.105.155201} {\bibfield  {journal} {\bibinfo  {journal} {Physical Review B}\ }\textbf {\bibinfo {volume} {105}},\ \bibinfo {pages} {155201} (\bibinfo {year} {2022})}\BibitemShut {NoStop}%
\bibitem [{\citenamefont {Freysoldt}\ \emph {et~al.}(2014)\citenamefont {Freysoldt}, \citenamefont {Grabowski}, \citenamefont {Hickel}, \citenamefont {Neugebauer}, \citenamefont {Kresse}, \citenamefont {Janotti},\ and\ \citenamefont {Van~de Walle}}]{34-freysoldt2014first-First-principles-calculations-for-point-defects-in-solids}%
  \BibitemOpen
  \bibfield  {author} {\bibinfo {author} {\bibfnamefont {C.}~\bibnamefont {Freysoldt}}, \bibinfo {author} {\bibfnamefont {B.}~\bibnamefont {Grabowski}}, \bibinfo {author} {\bibfnamefont {T.}~\bibnamefont {Hickel}}, \bibinfo {author} {\bibfnamefont {J.}~\bibnamefont {Neugebauer}}, \bibinfo {author} {\bibfnamefont {G.}~\bibnamefont {Kresse}}, \bibinfo {author} {\bibfnamefont {A.}~\bibnamefont {Janotti}}, \ and\ \bibinfo {author} {\bibfnamefont {C.~G.}\ \bibnamefont {Van~de Walle}},\ }\href {https://link.aps.org/doi/10.1103/RevModPhys.86.253} {\bibfield  {journal} {\bibinfo  {journal} {Reviews of Modern Physics}\ }\textbf {\bibinfo {volume} {86}},\ \bibinfo {pages} {253} (\bibinfo {year} {2014})}\BibitemShut {NoStop}%
\bibitem [{\citenamefont {Zhang}\ \emph {et~al.}(2020)\citenamefont {Zhang}, \citenamefont {Turiansky}, \citenamefont {Shen},\ and\ \citenamefont {Van~de Walle}}]{24-socIodineinterstitials}%
  \BibitemOpen
  \bibfield  {author} {\bibinfo {author} {\bibfnamefont {X.}~\bibnamefont {Zhang}}, \bibinfo {author} {\bibfnamefont {M.~E.}\ \bibnamefont {Turiansky}}, \bibinfo {author} {\bibfnamefont {J.-X.}\ \bibnamefont {Shen}}, \ and\ \bibinfo {author} {\bibfnamefont {C.~G.}\ \bibnamefont {Van~de Walle}},\ }\href {https://journals.aps.org/prb/abstract/10.1103/PhysRevB.101.140101} {\bibfield  {journal} {\bibinfo  {journal} {Physical Review B}\ }\textbf {\bibinfo {volume} {101}},\ \bibinfo {pages} {140101} (\bibinfo {year} {2020})}\BibitemShut {NoStop}%
\bibitem [{\citenamefont {Zhang}\ \emph {et~al.}(2022)\citenamefont {Zhang}, \citenamefont {Turiansky}, \citenamefont {Shen},\ and\ \citenamefont {Van~de Walle}}]{25-zhang2022defect}%
  \BibitemOpen
  \bibfield  {author} {\bibinfo {author} {\bibfnamefont {X.}~\bibnamefont {Zhang}}, \bibinfo {author} {\bibfnamefont {M.~E.}\ \bibnamefont {Turiansky}}, \bibinfo {author} {\bibfnamefont {J.-X.}\ \bibnamefont {Shen}}, \ and\ \bibinfo {author} {\bibfnamefont {C.~G.}\ \bibnamefont {Van~de Walle}},\ }\href {https://pubs.aip.org/aip/jap/article/131/9/090901/2836495/Defect-tolerance-in-halide-perovskites-A-first} {\bibfield  {journal} {\bibinfo  {journal} {Journal of Applied Physics}\ }\textbf {\bibinfo {volume} {131}},\ \bibinfo {pages} {090901} (\bibinfo {year} {2022})}\BibitemShut {NoStop}%
\bibitem [{\citenamefont {Zhang}\ and\ \citenamefont {Zhong}(2022)}]{35-zhang2022origins-Originsofp-Dopingincssni3}%
  \BibitemOpen
  \bibfield  {author} {\bibinfo {author} {\bibfnamefont {J.}~\bibnamefont {Zhang}}\ and\ \bibinfo {author} {\bibfnamefont {Y.}~\bibnamefont {Zhong}},\ }\href {https://onlinelibrary.wiley.com/doi/full/10.1002/ange.202212002} {\bibfield  {journal} {\bibinfo  {journal} {Angewandte Chemie}\ }\textbf {\bibinfo {volume} {134}},\ \bibinfo {pages} {e202212002} (\bibinfo {year} {2022})}\BibitemShut {NoStop}%
\bibitem [{\citenamefont {Sharma}\ \emph {et~al.}(2020)\citenamefont {Sharma}, \citenamefont {Kumar}, \citenamefont {Dev},\ and\ \citenamefont {Pilania}}]{26-predictml3predictabo3}%
  \BibitemOpen
  \bibfield  {author} {\bibinfo {author} {\bibfnamefont {V.}~\bibnamefont {Sharma}}, \bibinfo {author} {\bibfnamefont {P.}~\bibnamefont {Kumar}}, \bibinfo {author} {\bibfnamefont {P.}~\bibnamefont {Dev}}, \ and\ \bibinfo {author} {\bibfnamefont {G.}~\bibnamefont {Pilania}},\ }\href {https://pubs.aip.org/aip/jap/article/128/3/034902/1025656/Machine-learning-substitutional-defect-formation} {\bibfield  {journal} {\bibinfo  {journal} {Journal of Applied Physics}\ }\textbf {\bibinfo {volume} {128}},\ \bibinfo {pages} {034902} (\bibinfo {year} {2020})}\BibitemShut {NoStop}%
\bibitem [{\citenamefont {Mannodi-Kanakkithodi}\ and\ \citenamefont {Chan}(2022)}]{25-predictml2impuritiesinhalideperovskites}%
  \BibitemOpen
  \bibfield  {author} {\bibinfo {author} {\bibfnamefont {A.}~\bibnamefont {Mannodi-Kanakkithodi}}\ and\ \bibinfo {author} {\bibfnamefont {M.~K.}\ \bibnamefont {Chan}},\ }\href {https://link.springer.com/article/10.1007/s10853-022-06998-z} {\bibfield  {journal} {\bibinfo  {journal} {Journal of Materials Science}\ }\textbf {\bibinfo {volume} {57}},\ \bibinfo {pages} {10736} (\bibinfo {year} {2022})}\BibitemShut {NoStop}%
\bibitem [{\citenamefont {Mannodi-Kanakkithodi}\ \emph {et~al.}(2022)\citenamefont {Mannodi-Kanakkithodi}, \citenamefont {Xiang}, \citenamefont {Jacoby}, \citenamefont {Biegaj}, \citenamefont {Dunham}, \citenamefont {Gamelin},\ and\ \citenamefont {Chan}}]{24-predictmlzbuniversalmachinelearning}%
  \BibitemOpen
  \bibfield  {author} {\bibinfo {author} {\bibfnamefont {A.}~\bibnamefont {Mannodi-Kanakkithodi}}, \bibinfo {author} {\bibfnamefont {X.}~\bibnamefont {Xiang}}, \bibinfo {author} {\bibfnamefont {L.}~\bibnamefont {Jacoby}}, \bibinfo {author} {\bibfnamefont {R.}~\bibnamefont {Biegaj}}, \bibinfo {author} {\bibfnamefont {S.~T.}\ \bibnamefont {Dunham}}, \bibinfo {author} {\bibfnamefont {D.~R.}\ \bibnamefont {Gamelin}}, \ and\ \bibinfo {author} {\bibfnamefont {M.~K.}\ \bibnamefont {Chan}},\ }\href {https://www.sciencedirect.com/science/article/pii/S266638992200023X?via%3Dihub} {\bibfield  {journal} {\bibinfo  {journal} {Patterns}\ }\textbf {\bibinfo {volume} {3}},\ \bibinfo {pages} {100450} (\bibinfo {year} {2022})}\BibitemShut {NoStop}%
\bibitem [{\citenamefont {Varley}\ \emph {et~al.}(2017)\citenamefont {Varley}, \citenamefont {Samanta},\ and\ \citenamefont {Lordi}}]{varley2017descriptor}%
  \BibitemOpen
  \bibfield  {author} {\bibinfo {author} {\bibfnamefont {J.~B.}\ \bibnamefont {Varley}}, \bibinfo {author} {\bibfnamefont {A.}~\bibnamefont {Samanta}}, \ and\ \bibinfo {author} {\bibfnamefont {V.}~\bibnamefont {Lordi}},\ }\href {https://pubs.acs.org/doi/abs/10.1021/acs.jpclett.7b02333} {\bibfield  {journal} {\bibinfo  {journal} {The Journal of Physical Chemistry Letters}\ }\textbf {\bibinfo {volume} {8}},\ \bibinfo {pages} {5059} (\bibinfo {year} {2017})}\BibitemShut {NoStop}%
\bibitem [{\citenamefont {Kresse}\ and\ \citenamefont {Furthm{\"u}ller}(1996)}]{29-vasp}%
  \BibitemOpen
  \bibfield  {author} {\bibinfo {author} {\bibfnamefont {G.}~\bibnamefont {Kresse}}\ and\ \bibinfo {author} {\bibfnamefont {J.}~\bibnamefont {Furthm{\"u}ller}},\ }\href {https://journals.aps.org/prb/abstract/10.1103/PhysRevB.54.11169} {\bibfield  {journal} {\bibinfo  {journal} {Physical Review B}\ }\textbf {\bibinfo {volume} {54}},\ \bibinfo {pages} {11169} (\bibinfo {year} {1996})}\BibitemShut {NoStop}%
\bibitem [{\citenamefont {Bl{\"o}chl}(1994)}]{30-pseudoProjectoraugmented-wavemethod}%
  \BibitemOpen
  \bibfield  {author} {\bibinfo {author} {\bibfnamefont {P.~E.}\ \bibnamefont {Bl{\"o}chl}},\ }\href {https://journals.aps.org/prb/abstract/10.1103/PhysRevB.50.17953} {\bibfield  {journal} {\bibinfo  {journal} {Physical Review B}\ }\textbf {\bibinfo {volume} {50}},\ \bibinfo {pages} {17953} (\bibinfo {year} {1994})}\BibitemShut {NoStop}%
\bibitem [{\citenamefont {Heyd}\ \emph {et~al.}(2003)\citenamefont {Heyd}, \citenamefont {Scuseria},\ and\ \citenamefont {Ernzerhof}}]{31-Hybridfunctionals-based-on-screened-Coulomb-potential}%
  \BibitemOpen
  \bibfield  {author} {\bibinfo {author} {\bibfnamefont {J.}~\bibnamefont {Heyd}}, \bibinfo {author} {\bibfnamefont {G.~E.}\ \bibnamefont {Scuseria}}, \ and\ \bibinfo {author} {\bibfnamefont {M.}~\bibnamefont {Ernzerhof}},\ }\href {https://pubs.aip.org/aip/jcp/article/118/18/8207/460359/Hybrid-functionals-based-on-a-screened-Coulomb} {\bibfield  {journal} {\bibinfo  {journal} {The Journal of Chemical Physics}\ }\textbf {\bibinfo {volume} {118}},\ \bibinfo {pages} {8207} (\bibinfo {year} {2003})}\BibitemShut {NoStop}%
\bibitem [{\citenamefont {Chen}\ \emph {et~al.}(2012)\citenamefont {Chen}, \citenamefont {Yu}, \citenamefont {Shum}, \citenamefont {Wang}, \citenamefont {Pfenninger}, \citenamefont {Vockic}, \citenamefont {Midgley},\ and\ \citenamefont {Kenney}}]{32-egcssni3}%
  \BibitemOpen
  \bibfield  {author} {\bibinfo {author} {\bibfnamefont {Z.}~\bibnamefont {Chen}}, \bibinfo {author} {\bibfnamefont {C.}~\bibnamefont {Yu}}, \bibinfo {author} {\bibfnamefont {K.}~\bibnamefont {Shum}}, \bibinfo {author} {\bibfnamefont {J.~J.}\ \bibnamefont {Wang}}, \bibinfo {author} {\bibfnamefont {W.}~\bibnamefont {Pfenninger}}, \bibinfo {author} {\bibfnamefont {N.}~\bibnamefont {Vockic}}, \bibinfo {author} {\bibfnamefont {J.}~\bibnamefont {Midgley}}, \ and\ \bibinfo {author} {\bibfnamefont {J.~T.}\ \bibnamefont {Kenney}},\ }\href {https://www.sciencedirect.com/science/article/pii/S002223131100500X} {\bibfield  {journal} {\bibinfo  {journal} {Journal of Luminescence}\ }\textbf {\bibinfo {volume} {132}},\ \bibinfo {pages} {345} (\bibinfo {year} {2012})}\BibitemShut {NoStop}%
\bibitem [{\citenamefont {Freysoldt}\ \emph {et~al.}(2009)\citenamefont {Freysoldt}, \citenamefont {Neugebauer},\ and\ \citenamefont {Van~de Walle}}]{22-freysoldt2009fully}%
  \BibitemOpen
  \bibfield  {author} {\bibinfo {author} {\bibfnamefont {C.}~\bibnamefont {Freysoldt}}, \bibinfo {author} {\bibfnamefont {J.}~\bibnamefont {Neugebauer}}, \ and\ \bibinfo {author} {\bibfnamefont {C.~G.}\ \bibnamefont {Van~de Walle}},\ }\href {https://journals.aps.org/prl/abstract/10.1103/PhysRevLett.102.016402} {\bibfield  {journal} {\bibinfo  {journal} {Physical Review Letters}\ }\textbf {\bibinfo {volume} {102}},\ \bibinfo {pages} {016402} (\bibinfo {year} {2009})}\BibitemShut {NoStop}%
\bibitem [{\citenamefont {Irham}\ \emph {et~al.}(2022)\citenamefont {Irham}, \citenamefont {Tejo~Baskoro}, \citenamefont {Permatasari},\ and\ \citenamefont {Iskandar}}]{18-irham2022toward}%
  \BibitemOpen
  \bibfield  {author} {\bibinfo {author} {\bibfnamefont {M.~A.}\ \bibnamefont {Irham}}, \bibinfo {author} {\bibfnamefont {F.~H.}\ \bibnamefont {Tejo~Baskoro}}, \bibinfo {author} {\bibfnamefont {F.~A.}\ \bibnamefont {Permatasari}}, \ and\ \bibinfo {author} {\bibfnamefont {F.}~\bibnamefont {Iskandar}},\ }\href {https://pubs.acs.org/doi/abs/10.1021/acs.jpcc.1c10315} {\bibfield  {journal} {\bibinfo  {journal} {The Journal of Physical Chemistry C}\ }\textbf {\bibinfo {volume} {126}},\ \bibinfo {pages} {5256} (\bibinfo {year} {2022})}\BibitemShut {NoStop}%
\bibitem [{\citenamefont {Li}\ \emph {et~al.}(2024)\citenamefont {Li}, \citenamefont {Khamdang},\ and\ \citenamefont {Wang}}]{li2024cssni3_surface_dft}%
  \BibitemOpen
  \bibfield  {author} {\bibinfo {author} {\bibfnamefont {K.}~\bibnamefont {Li}}, \bibinfo {author} {\bibfnamefont {C.}~\bibnamefont {Khamdang}}, \ and\ \bibinfo {author} {\bibfnamefont {M.}~\bibnamefont {Wang}},\ }\href {https://journals.aps.org/prmaterials/abstract/10.1103/PhysRevMaterials.8.093401} {\bibfield  {journal} {\bibinfo  {journal} {Phys. Rev. Mater.}\ }\textbf {\bibinfo {volume} {8}},\ \bibinfo {pages} {093401} (\bibinfo {year} {2024})}\BibitemShut {NoStop}%
\bibitem [{\citenamefont {Zhang}\ \emph {et~al.}(2023)\citenamefont {Zhang}, \citenamefont {Zhang}, \citenamefont {Turiansky},\ and\ \citenamefont {Van~de Walle}}]{23-zhang2023iodine}%
  \BibitemOpen
  \bibfield  {author} {\bibinfo {author} {\bibfnamefont {J.}~\bibnamefont {Zhang}}, \bibinfo {author} {\bibfnamefont {X.}~\bibnamefont {Zhang}}, \bibinfo {author} {\bibfnamefont {M.~E.}\ \bibnamefont {Turiansky}}, \ and\ \bibinfo {author} {\bibfnamefont {C.~G.}\ \bibnamefont {Van~de Walle}},\ }\href {https://journals.aps.org/prxenergy/abstract/10.1103/PRXEnergy.2.013008} {\bibfield  {journal} {\bibinfo  {journal} {PRX Energy}\ }\textbf {\bibinfo {volume} {2}},\ \bibinfo {pages} {013008} (\bibinfo {year} {2023})}\BibitemShut {NoStop}%
\bibitem [{\citenamefont {Goldschmidt}(1926)}]{goldschmidt1926gesetze}%
  \BibitemOpen
  \bibfield  {author} {\bibinfo {author} {\bibfnamefont {V.~M.}\ \bibnamefont {Goldschmidt}},\ }\href {https://link.springer.com/article/10.1007/bf01507527} {\bibfield  {journal} {\bibinfo  {journal} {Naturwissenschaften}\ }\textbf {\bibinfo {volume} {14}},\ \bibinfo {pages} {477} (\bibinfo {year} {1926})}\BibitemShut {NoStop}%
\bibitem [{\citenamefont {Li}\ \emph {et~al.}(2008)\citenamefont {Li}, \citenamefont {Lu}, \citenamefont {Ding}, \citenamefont {Feng}, \citenamefont {Gao},\ and\ \citenamefont {Guo}}]{li2008formability}%
  \BibitemOpen
  \bibfield  {author} {\bibinfo {author} {\bibfnamefont {C.}~\bibnamefont {Li}}, \bibinfo {author} {\bibfnamefont {X.}~\bibnamefont {Lu}}, \bibinfo {author} {\bibfnamefont {W.}~\bibnamefont {Ding}}, \bibinfo {author} {\bibfnamefont {L.}~\bibnamefont {Feng}}, \bibinfo {author} {\bibfnamefont {Y.}~\bibnamefont {Gao}}, \ and\ \bibinfo {author} {\bibfnamefont {Z.}~\bibnamefont {Guo}},\ }\href {https://onlinelibrary.wiley.com/iucr/doi/10.1107/S0108768108032734} {\bibfield  {journal} {\bibinfo  {journal} {Acta Crystallographica Section B: Structural Science}\ }\textbf {\bibinfo {volume} {64}},\ \bibinfo {pages} {702} (\bibinfo {year} {2008})}\BibitemShut {NoStop}%
\bibitem [{\citenamefont {Shannon}(1976)}]{shannon1976IR}%
  \BibitemOpen
  \bibfield  {author} {\bibinfo {author} {\bibfnamefont {R.~D.}\ \bibnamefont {Shannon}},\ }\href {https://journals.iucr.org/paper?pii=S0567739476001551} {\bibfield  {journal} {\bibinfo  {journal} {Foundations of Crystallography}\ }\textbf {\bibinfo {volume} {32}},\ \bibinfo {pages} {751} (\bibinfo {year} {1976})}\BibitemShut {NoStop}%
\bibitem [{\citenamefont {Cohen}\ \emph {et~al.}(2009)\citenamefont {Cohen}, \citenamefont {Huang}, \citenamefont {Chen}, \citenamefont {Benesty}, \citenamefont {Benesty}, \citenamefont {Chen}, \citenamefont {Huang},\ and\ \citenamefont {Cohen}}]{cohen2009pearson}%
  \BibitemOpen
  \bibfield  {author} {\bibinfo {author} {\bibfnamefont {I.}~\bibnamefont {Cohen}}, \bibinfo {author} {\bibfnamefont {Y.}~\bibnamefont {Huang}}, \bibinfo {author} {\bibfnamefont {J.}~\bibnamefont {Chen}}, \bibinfo {author} {\bibfnamefont {J.}~\bibnamefont {Benesty}}, \bibinfo {author} {\bibfnamefont {J.}~\bibnamefont {Benesty}}, \bibinfo {author} {\bibfnamefont {J.}~\bibnamefont {Chen}}, \bibinfo {author} {\bibfnamefont {Y.}~\bibnamefont {Huang}}, \ and\ \bibinfo {author} {\bibfnamefont {I.}~\bibnamefont {Cohen}},\ }\href {https://doi.org/10.1007/978-3-642-00296-0_5} {\bibfield  {journal} {\bibinfo  {journal} {Noise Reduction in Speech Processing}\ ,\ \bibinfo {pages} {1}} (\bibinfo {year} {2009})}\BibitemShut {NoStop}%
\bibitem [{\citenamefont {Pedregosa}(2011)}]{pedregosa2011scikit}%
  \BibitemOpen
  \bibfield  {author} {\bibinfo {author} {\bibfnamefont {F.}~\bibnamefont {Pedregosa}},\ }\href {https://jmlr.csail.mit.edu/papers/v12/pedregosa11a.html} {\bibfield  {journal} {\bibinfo  {journal} {Journal of Machine Learning Research}\ }\textbf {\bibinfo {volume} {12}},\ \bibinfo {pages} {2825} (\bibinfo {year} {2011})}\BibitemShut {NoStop}%
\bibitem [{\citenamefont {Seeger}(2004)}]{gaussian}%
  \BibitemOpen
  \bibfield  {author} {\bibinfo {author} {\bibfnamefont {M.}~\bibnamefont {Seeger}},\ }\href {https://pubmed.ncbi.nlm.nih.gov/15112367/} {\bibfield  {journal} {\bibinfo  {journal} {International Journal of Neural Systems}\ }\textbf {\bibinfo {volume} {14}},\ \bibinfo {pages} {69} (\bibinfo {year} {2004})}\BibitemShut {NoStop}%
\bibitem [{\citenamefont {Puga}\ \emph {et~al.}(2015)\citenamefont {Puga}, \citenamefont {Krzywinski},\ and\ \citenamefont {Altman}}]{bay}%
  \BibitemOpen
  \bibfield  {author} {\bibinfo {author} {\bibfnamefont {J.~L.}\ \bibnamefont {Puga}}, \bibinfo {author} {\bibfnamefont {M.}~\bibnamefont {Krzywinski}}, \ and\ \bibinfo {author} {\bibfnamefont {N.}~\bibnamefont {Altman}},\ }\href {https://www.nature.com/articles/nmeth.3335} {\bibfield  {journal} {\bibinfo  {journal} {Nature Methods}\ }\textbf {\bibinfo {volume} {12}},\ \bibinfo {pages} {277} (\bibinfo {year} {2015})}\BibitemShut {NoStop}%
\bibitem [{\citenamefont {Duvenaud}(2014)}]{duvenaud2014automatic}%
  \BibitemOpen
  \bibfield  {author} {\bibinfo {author} {\bibfnamefont {D.}~\bibnamefont {Duvenaud}},\ }\emph {\bibinfo {title} {Automatic model construction with Gaussian processes}},\ \href {https://www.repository.cam.ac.uk/items/a54f0711-f777-4ee1-a685-3b76dae5ad91} {Ph.D. thesis} (\bibinfo {year} {2014})\BibitemShut {NoStop}%
\bibitem [{\citenamefont {Vovk}(2013)}]{vovk2013kernel}%
  \BibitemOpen
  \bibfield  {author} {\bibinfo {author} {\bibfnamefont {V.}~\bibnamefont {Vovk}},\ }in\ \href {https://link.springer.com/chapter/10.1007/978-3-642-41136-6_11} {\emph {\bibinfo {booktitle} {Empirical inference: Festschrift in honor of vladimir n. vapnik}}}\ (\bibinfo  {publisher} {Springer},\ \bibinfo {year} {2013})\ pp.\ \bibinfo {pages} {105--116}\BibitemShut {NoStop}%
\bibitem [{\citenamefont {Breiman}(2001)}]{rfr}%
  \BibitemOpen
  \bibfield  {author} {\bibinfo {author} {\bibfnamefont {L.}~\bibnamefont {Breiman}},\ }\href {https://link.springer.com/article/10.1023/a:1010933404324} {\bibfield  {journal} {\bibinfo  {journal} {Machine Learning}\ }\textbf {\bibinfo {volume} {45}},\ \bibinfo {pages} {5} (\bibinfo {year} {2001})}\BibitemShut {NoStop}%
\end{thebibliography}%

\end{document}